\makeatletter
\newcommand{\dontusepackage}[2][]{%
  \@namedef{ver@#2.sty}{9999/12/31}%
  \@namedef{opt@#2.sty}{#1}}
\makeatother
\dontusepackage{subfigure}

\documentclass[]{article}

\usepackage{lmodern}
\usepackage{amssymb,amsmath}
\usepackage{ifxetex,ifluatex}
\usepackage[usenames,dvipsnames]{color}
\usepackage{fixltx2e} 
\ifnum 0\ifxetex 1\fi\ifluatex 1\fi=0 
  \usepackage[T1]{fontenc}
  \usepackage[utf8]{inputenc}
\else 
  \ifxetex
    \usepackage{mathspec}
    \usepackage{xltxtra,xunicode}
  \else
    \usepackage{fontspec}
  \fi
  \defaultfontfeatures{Mapping=tex-text,Scale=MatchLowercase}
  
\fi
\IfFileExists{upquote.sty}{\usepackage{upquote}}{}
\IfFileExists{microtype.sty}{%
\usepackage{microtype}
\UseMicrotypeSet[protrusion]{basicmath} 
}{}
\usepackage[margin=1.0in,bottom=1.5in]{geometry}
\usepackage[square,numbers,sort&compress]{natbib}
\usepackage{listings}
\usepackage{graphicx}
\makeatletter
\def\maxwidth{\ifdim\Gin@nat@width>\linewidth\linewidth\else\Gin@nat@width\fi}
\def\maxheight{\ifdim\Gin@nat@height>\textheight\textheight\else\Gin@nat@height\fi}
\makeatother
\setkeys{Gin}{width=\maxwidth,height=\maxheight,keepaspectratio}
\usepackage{caption}
\usepackage{float}
\usepackage{subfig}
\usepackage{algorithm} 
\let\scholmdAlgorithm\algorithm
\let\endscholmdAlgorithm\endalgorithm
\let\algorithm\relax \let\endalgorithm\relax

{
 \catcode`\*=11\relax
 \global\let\scholmdAlgorithm*\algorithm*
 \global\let\endscholmdAlgorithm*\endalgorithm*
 \global\let\algorithm*\relax 
 \global\let\endalgorithm*\relax
}
\ifxetex
  \usepackage[setpagesize=false, 
              unicode=false, 
              xetex]{hyperref}
\else
  \usepackage[unicode=true]{hyperref}
\fi
\hypersetup{breaklinks=true,
            bookmarks=true,
            pdfauthor={},
            pdftitle={Deep Bayesian inference for seismic imaging with tasks},
            colorlinks=true,
            citecolor=black,
            urlcolor=blue,
            linkcolor=black,
            pdfborder={0 0 0}}
\usepackage[preprint]{neurips_2019}
\usepackage{url}
\usepackage{booktabs}
\usepackage{amsfonts}
\usepackage{nicefrac}
\usepackage{microtype}

\newcommand{\R}{\mbox{$\mathbb{R}$}}
\newcommand{\B}{\mathbf}

\title{Deep Bayesian inference for seismic imaging with tasks}
\author{Ali Siahkoohi\\School of Computational Science and Engineering,\\Georgia
Institute of Technology\\\texttt{alisk@gatech.edu}\\\And
Gabrio Rizzuti\\Georgia Institute of Technology\\Now at Utrecht
University\\\texttt{g.rizzuti@uu.nl}\\\And
Felix J. Herrmann\\School of Computational Science and
Engineering,\\Georgia Institute of
Technology\\\texttt{felix.herrmann@gatech.edu}}
\date{}

\begin{document}
\maketitle
\begin{abstract}
We propose to use techniques from Bayesian inference and deep neural
networks to translate uncertainty in seismic imaging to uncertainty in
tasks performed on the image, such as horizon tracking. Seismic imaging
is an ill-posed inverse problem because of bandwidth and aperture
limitations, which is hampered by the presence of noise and
linearization errors. Many regularization methods, such as
transform-domain sparsity promotion, have been designed to deal with the
adverse effects of these errors, however, these methods run the risk of
biasing the solution and do not provide information on uncertainty in
the image space and how this uncertainty impacts certain tasks on the
image. A systematic approach is proposed to translate uncertainty due to
noise in the data to confidence intervals of automatically tracked
horizons in the image. The uncertainty is characterized by a
convolutional neural network (CNN) and to assess these uncertainties,
samples are drawn from the posterior distribution of the CNN weights,
used to parameterize the image. Compared to traditional priors, it is
argued in the literature that these CNNs introduce a flexible inductive
bias that is a surprisingly good fit for a diverse set of problems,
including medical imaging, compressive sensing, and diffraction
tomography. The method of stochastic gradient Langevin dynamics is
employed to sample from the posterior distribution. This method is
designed to handle large scale Bayesian inference problems with
computationally expensive forward operators as in seismic imaging. Aside
from offering a robust alternative to maximum a posteriori estimate that
is prone to overfitting, access to these samples allow us to translate
uncertainty in the image, due to noise in the data, to uncertainty on
the tracked horizons. For instance, it admits estimates for the
pointwise standard deviation on the image and for confidence intervals
on its automatically tracked horizons.
\end{abstract}

\section{Introduction}\label{introduction}

Due to the presence of shadow zones, coherent linearization errors, and
noisy finite-aperture measured data, seismic imaging involves an
ill-conditioned linear inverse problem
\citep{lambare1992iterative, schuster1993least, nemeth1999least}.
Relying on a single estimate for the model may be subject to data
overfit \citep{malinverno2004expanded} and negatively impacts the
quality of the obtained seismic image and tasks performed on it. Casting
the seismic imaging problem into a probabilistic framework allows for a
more comprehensive description of its solution space
\citep{tarantola2005inverse}. The ``solution'' of the inverse problem is
then a probability distribution over the model space and is commonly
referred to as the posterior distribution.

Aside from the computational challenges associated with uncertainty
quantification (UQ) in geophysical inverse problems
\citep{malinverno2004expanded, tarantola2005inverse, malinverno2006two, MartinMcMC2012, chevron2017, fang2018uqfip, stuart2019two, zhao2019gradient, kotsi2020},
the choice of prior distributions in Bayesian frameworks is crucial.
Recent attempts mostly rely on handcrafted priors, i.e., priors chosen
solely based on their simplicity and applicability. For example,
restricting feasible solutions to layered media with specific
orientations
\citep{doi:10.1046/j.1365-246X.2002.01847.x, zhu2018seismic, visser2019bayesian},
or satisfying regularity conditions related to model parameters or
derivatives thereof
\citep{malinverno2004expanded, malinverno2006two, vanLeeuwen2010IJGswi, herrmann11GPelsqIm, MartinMcMC2012, lu2015l1, tu2014fis, zhu2016bayesian, Ely2018, fang2018uqfip, stuart2019two, zhao2019gradient, izzatullah2020, kotsi2020}.
While effective in controlled settings, handcrafted priors might
introduce unwanted bias to the solution. Recent deep-learning based
approaches
\citep{mosser2018stochastic, zhang2019regularized, fang2020deep, Zhilongsegam2020-3428013.1, liu2020deep, Mossereage2020, sun2020extrapolated, Youzuo2020, kazei2021mapping, kothari2021trumpets, kumar2021SEGeuq, siahkoohi2021Seglbe},
on the other hand, learn a prior distribution from available
data\footnote{We use the word ``data'' interchangeably. In the context
  of inverse problems ``data'' refers to observed data. In the context
  of machine learning data-driven priors refer to priors derived from
  available samples from an unknown distribution. These are also
  commonly referred to as ``data''. The meaning of the word ``data''
  will be clear from the context.}. While certainly providing a better
description of the available prior information when compared to generic
handcrafted priors, they may affect the outcome of Bayesian inference
more seriously when out-of-distribution data is considered, e.g., when
the training data is not fully representative of a given scenario.
Unfortunately, unlike deep-learning based inversion approaches in other
imaging modalities, e.g., medical imaging
\citep{adler2018deep, welling2019, pmlr-v119-asim20a, hauptmann2020deep, sriram2020end, mukherjee2021end},
we generally do not have access to high-fidelity information about the
Earth's subsurface. This, together with the Earth's strong heterogeneity
across geological scenarios, might limit the scope of data-driven
approaches that heavily rely on pretraining
\citep{siahkoohi2019transfer, kaur2020improving, Willett2020, Youzuo2020_physics_consistent, sun2020elastic, zhang2020SEGtli, barbano2021unsupervised, siahkoohi2020ABIpto, qu2021training, vrolijk2021source}.

In this work, we take advantage of a novel prior recently deployed in
computer vision and geophysics
\citep{Lempitsky, arridge_oktem_2019, Cheng_2019_CVPR, gadelha2019shape, liu2019deep, wu2019parametric, dittmer2020regularization, Fantong2020, shi2020deep, siahkoohi2020EAGEdlb, siahkoohi2020SEGuqi, siahkoohi2020SEGwdp, tolle2021mean},
known as the deep prior, which utilizes the inductive bias
\citep{mitchell1980need} of untrained convolutional neural networks
(CNNs) as a prior. This approach is tantamount to restricting feasible
models to the range of an untrained CNN with a fixed input and randomly
initialized weights. Via this reparameterization the weights of the CNN
become the new unknowns in seismic imaging and this change of variable
leads to a ``prior'' on the image space that excludes noisy artifacts,
as long as overfitting is prevented \citep{Lempitsky}. This has the
potential benefit of being less restrictive than handcrafted priors
while not needing training data as approaches based on using pretrained
networks \citep{Lempitsky}. To formally cast the deep prior into a
Bayesian framework, we impose a Gaussian distribution on the CNN
weights, which is a common regularization strategy in training deep CNNs
\citep{krogh1992simple, Goodfellow-et-al-2016}. To perform uncertainty
quantification for seismic imaging, we sample from the posterior
distribution of the CNN weights by running preconditioned stochastic
gradient Langevin dynamics
\citep[SGLD,][]{welling2011bayesian, li2016preconditioned}, a
gradient-based Markov chain Monte Carlo (MCMC) sampling method developed
for Bayesian inference of deep CNNs with large training datasets.

A crucial objective of our study is translating the uncertainty in
seismic imaging to uncertainty in downstream tasks such as horizon
tracking, semantic segmentation, and tracking CO$_2$ plumes in carbon
capture and sequestration projects. Horizon tracking, which this papers
focuses on, is a task performed after imaging that leads to a
stratigraphic model. Horizon trackers use well data and seismic images
to delineate stratigraphy and are typically sensitive to structural and
stratigraphic unconformities. In these challenging areas, the horizons
do not continuously extend spatially, e.g., may be discontinuous due to
vertical displacement via faults, hence tracking horizons across
unconformities may not be trivial. Failure to include uncertainty on
tracked horizons have major implications on the identification of risk.
Since the accuracy of horizon tracking is directly linked to the quality
of the seismic image, we systematically incorporate uncertainties of
seismic imaging into horizon tracking. We achieve this by feeding
samples from the imaging posterior to an automatic horizon tracker
\citep{wu2018least} and obtain an ensemble of likely horizons in the
image. Compared to conventional imaging and manual tracking of horizons,
our approach allows us to rigorously quantify uncertainty in the
location of the horizons due to noise in shot records and modeling
errors, e.g., linearization errors. Our probabilistic framework also
admits nondeterministic horizon trackers, e.g., uncertain control points
or multiple human interpreters. There are parallels between the
probabilistic framework we propose for quantifying uncertainty in
downstream tasks and the interrogation theory \citep{interrogation}. The
purpose of this theory is to answer questions about an unknown quantity
by designing experiments (inverse problems) that facilitate answering
the question. The probabilistic framework we developed can be described
as an application of interrogation theory in that the seismic survey and
shot records are provided with no need to design further experiments,
and the question involves quantifying uncertainty in horizon tracking.
Our probabilistic framework differs fundamentally from other recently
developed automatic seismic horizon trackers based on machine learning
\citep[see
e.g.,][]{peters2019, geng2020deep, peters2020eage, shi2020waveform}
because horizon uncertainty is ultimately driven by data (through the
intermediate imaging distribution), and not from label (control point)
uncertainty alone.

In the following sections, we first mathematically formulate deep-prior
based seismic imaging, by introducing the likelihood function and the
deep prior approach. Next, we describe our proposed SGLD-based sampling
approach and its challenges. Subsequently, we introduce a framework to
tie uncertainties in imaging to uncertainties in horizon tracking, which
allows for deterministic and nondeterministic horizon tracking. We
present two realistic examples derived from real seismic image volumes
obtained in different geological settings. These numerical experiments
are designed to showcase the ability of the proposed deep-prior based
approach to produce seismic images with limited artifacts. We conclude
by demonstrating our probabilistic horizon tracking approach, which
includes estimates for confidence intervals associated with the imaged
horizons in the two aforementioned examples.

\section{Theory}\label{theory}

The goal of this paper is to understand how errors in the data due to
noise and linearization assumptions affect the uncertainty of seismic
images and typical tasks carried out on these images. We begin with an
introduction of the linearized forward model, which forms the basis of
seismic imaging via reverse-time migration and discuss Bayesian imaging
with regularization via so-called deep priors.

\subsection{Seismic imaging}\label{seismic-imaging}

In its simplest acoustic form, reverse-time migration follows directly
from linearizing the acoustic wave equation around a known, slowly
varying background model---i.e., the spatial distribution of the squared
slowness. Traditionally, the process of seismic imaging is concerned
with estimating the short-wavelength components of the squared-slowness,
denoted by $\delta \B{m}$, from ${n_s}$ processed shot records collected
in the vector $\B{d} = \left \{\B{d}_{i}\right \}_{i=1}^{n_s}$. In most
cases, these indirect measurements are recorded along the surface or
ocean bottom with sources $\left \{\B{q}_{i}\right \}_{i=1}^{n_s}$
located at or near the surface. The placement of the sources and
receiver near the surface leads to more uncertainty in the deeper areas
of the image.

The unknown ground truth perturbation model $\delta \B{m}^{\ast}$ is
linearly related to the data via
\begin{equation}
\B{d}_i = \B{J}(\B{m}_0, \B{q}_i)
    \delta \B{m}^{\ast} + \boldsymbol{\epsilon}_i, \quad
    \boldsymbol{\epsilon}_i \sim \mathrm{N} (\B{0}, \sigma^2
    \B{I}),
\label{linear-fwd-op}
\end{equation}
 where $\B{J}(\B{m}_0, \B{q}_i)$ corresponds to the linearized Born
scattering operator for the $i\text{th}$ source and the background
squared slowness model $\B{m}_0$. Because of possible errors in the
processed data, the presence of noise, and linearization errors, the
above expression contains the noise term $\boldsymbol{\epsilon}_i$.
While other choices can be made, we assume this noise to be distributed
according to a zero-centered Gaussian distribution with known covariance
matrix $\sigma^2 \mathbf{I}$. For small $\sigma$ and a kinematically
correct background model $\B{m}_0$, the above linear relationship can be
inverted by minimizing
\begin{equation}
\min_{\delta \B{m}} \sum_{i=1}^{n_s} \big \|
    \B{d}_i- \B{J}(\B{m}_0, \B{q}_i)
    \delta \B{m} \big \|_2^2.
\label{imaging-opt}
\end{equation}
 While this approach is in principle capable of producing high-fidelity
true-amplitude images
\citep{valenciano2008imaging, dong2012least, zeng2014least}, the noise
term is in practice never negligible and may adversely affect the image
quality \citep{nemeth1999least} especially in situations where the
source spectrum is narrow band and the aperture limited. Therefore not
only the problem in equation~\ref{imaging-opt} requires regularization
but also calls for a statistical inference framework that allows us to
draw conclusions in the presence of uncertainty.

\subsection{Probabilistic imaging with Bayesian
inference}\label{probabilistic-imaging-with-bayesian-inference}

To account for uncertainties in the image induced by the random noise
term $\boldsymbol{\epsilon}_i$ in equation~\ref{linear-fwd-op}, we
follow the seminal work of \citet{tarantola2005inverse} and cast our
noisy imaging as a Bayesian inverse problem. Instead of calculating a
single image by solving equation~\ref{imaging-opt}, we assign
probabilities to a family of images that fit the observed data to
various degrees. This distribution is known as the posterior
distribution. In this Bayesian framework, the solution to the inverse
problem, i.e., the image, and the noise in the observed data are
considered random variables. According to Bayes' rule, the conditional
posterior distribution, denoted by $p_{\text{post}}$, states that
\begin{equation}
p_{\text{post}} \left (\delta \B{m} \mid \B{d} \right)
    \propto  p_{\text{like}} (\B{d} \mid \delta
    \B{m})\, p_{\text{prior}} (\delta \B{m}).
\label{bayes}
\end{equation}
 In this expression, $p_{\text{like}} $ is the likelihood function,
which is related to the probability density function (PDF) of the noise,
and $p_{\text{prior}}$ is the prior PDF of the image, which encodes
prior beliefs on the unknown perturbations $\delta\B{m}$. This prior
distribution assigns probabilities to all potential seismic images
before incorporating the data via the likelihood. The constant of
proportionality in equation~\ref{bayes} corresponds to the PDF of the
observed data, which is independent of $\delta \B{m}$. Based on the
distribution of the noise, the likelihood term measures how well the
forward modeled data (equation~\ref{linear-fwd-op}) and observed data
agree.

As stated by Bayes' rule, the posterior PDF of $\delta \B{m}$, denoted
by $p_{\text{post}} \left (\delta \B{m} \mid \B{d} \right)$, is
proportional to the product of the likelihood and the prior PDF, given
observed data. The log-likelihood function takes the following form:
\begin{equation}
\begin{aligned}
 - \log p_{\text{like}} \left ( \B{d}
    \mid \delta \B{m} \right ) & = -\sum_{i=1}^{n_s} \log p_{\text{like}}
    \left ( \B{d}_{i}\mid \delta \B{m} \right ) \\
& = \frac{1}{2 \sigma^2} \sum_{i=1}^{n_s}\big  \| \B{d}_i-
    \B{J}(\B{m}_0, \B{q}_i) \delta \B{m} \big \|_2^2 +
    \text{const},
\end{aligned}
\label{imaging-likelihood}
\end{equation}
 where the constant term is independent of $\delta \B{m}$. For the
uncorrelated Gaussian noise assumption, this negative log-likelihood
function equals the squared $\ell_2$-norm of the residual scaled by the
noise variance $\sigma^2$.

Aside from depending on the residual, i.e., the difference between
observed and modeled data, for each shot record, the choice of the prior
influences the posterior distribution. Before the advent of data-driven
methods involving generative neural networks, the definitions of priors
were mostly handcrafted and often based on somewhat ad hoc Gaussian or
Laplacian distributions in the physical or in some transformed domain
\citep{vanLeeuwen2010IJGswi, herrmann11GPelsqIm, lu2015l1, tu2014fis}.
While these approaches have proven to be useful and are theoretically
well understood \citep{donoho2006compressed}, there is always a risk of
a biased outcome something we would like to avoid. On the other hand,
using pretrained generative networks as priors has proven to be
effective
\citep{pmlr-v70-bora17a, mosser2018stochastic, pmlr-v119-asim20a, fang2020deep, siahkoohi2020ABIpto}.
However, their success hinges on the quality of pretraining and having
access to a fully representative training data that accurately captures
the prior distribution. Since we are dealing with highly complex
heterogeneity of the Earth subsurface to which we have limited access,
we will stay away from data-driven methods to train a neural network to
act as a prior.

\subsection{Deep priors}\label{deep-priors}

Following recent work by \citet{Lempitsky}, we avoid using the need to
have access to realizations of true perturbations by using untrained
generative CNNs as priors. We fix a random input latent variable and use
a randomly initialized \citep{pmlrv9glorot10a} CNN with a special
architecture \citep{Lempitsky} to reparameterize the unknown
perturbations $\delta\B{m}$ in terms of CNN weights. Given the shot
data, we minimize the data misfit with respect to the CNN weights on
which we impose a Gaussian prior. The CNN's architecture (see details in
Appendix A) and the Gaussian prior imposed on its weights act as a
regularization in the image space that avoids representing incoherent
noisy artifacts, as long as overfitting is prevented \citep{Lempitsky}.

To be more specific, let ${g} (\B{z}, \B{w})\in\R^{N}$, with the $N$ the
number of gridpoints in the image, denote a untrained, specially
designed, CNN \citep{Lempitsky} with fixed input
$\B{z} \sim \mathrm{N}( \B{0}, \B{I})$ with the same size as the image
and unknown weights $\B{w}\in\R^M$ with $M\gg N$. Restricting the
unknown perturbation model to the output of the CNN, i.e.,
$\delta \B{m} = {g} (\B{z}, \B{w})$, corresponds to a nonlinear
representation for the image and the following expression for the
likelihood function:
\begin{equation}
\begin{aligned}
 - \log p_{\text{like}} \left ( \B{d} \mid \B{w} \right )
&= -\sum_{i=1}^{n_s} \log p_{\text{like}}
    \left ( \B{d}_{i}\mid \B{w} \right ) \\
& =  \frac{1}{2 \sigma^2} \sum_{i=1}^{n_s} \big  \| \B{d}_i-
    \B{J}(\B{m}_0, \B{q}_i) g(\B{z},
    \B{w}) \big \|_2^2 + \text{const},
\end{aligned}
\label{deep-prior-likelihood}
\end{equation}
 with the constant term independent of $\B{w}$. In essence, deep priors
correspond to a nonlinear ``change of variables'' where the unknowns are
the CNN weights and the image is constrained to the range of the CNN
output for a fixed random input. Compared to data-driven methods, no
training samples are needed. While the nonlinearity makes it more
difficult to minimize the likelihood term (the likelihood in
equation~\ref{imaging-likelihood}), a zero-centered Gaussian prior for
the weights with covariance $\lambda^{-2}\B{I}$ suffices thanks to the
overparameterization of the CNN ($M\gg N$). With this Gaussian prior on
the weights, the posterior distribution for the weights given the data
reads
\begin{equation}
p_{\text{post}} \left ( \B{w} \mid \B{d} \right )
\propto \left [ \prod_{i=1}^{{n_s}} p_{\text{like}} \left (
    \B{d}_{i} \mid \B{w} \right ) \right ]
    \mathrm{N} \big (\B{w} \mid \B{0}, \lambda^{-2}\B{I} \big )
\label{deep-prior}
\end{equation}
 where $\mathrm{N} \big (\B{w} \mid \B{0}, \lambda^{-2}\B{I} \big )$
stands for the PDF of the zero-centered Gaussian prior. Given
equation~\ref{deep-prior-likelihood}, the negative log-posterior
distribution becomes
\begin{equation}
\begin{aligned}
 - \log p_{\text{post}} \left ( \B{w}
    \mid \B{d} \right )
& = - \left [ \sum_{i=1}^{{n_s}} \log p_{\text{like}} \left (
    \B{d}_{i}\mid\B{w} \right ) \right ] -
    \log \mathrm{N} \big (\B{w} \mid \B{0}, \lambda^{-2}\B{I} \big ) + \text{const} \\
& = \frac{1}{2 \sigma^2} \sum_{i=1}^{n_s}\big  \| \B{d}_i-
    \B{J}(\B{m}_0, \B{q}_i) {g} (\B{z},
    \B{w}) \big \|_2^2 + \frac{\lambda^2}{2}\big  \| \B{w} \big \|_2^2
    + \text{const}.
\end{aligned}
\label{imaging-obj}
\end{equation}
 Compared to conventional formulations of Bayesian inference, knowledge
of the deep prior resides both in the likelihood term, through the
reparameterization of the image as the output of a CNN, and in the
traditional $\lambda$ weighted $\ell_2$-norm squared term. This is
different from the traditional Bayesian settings where prior information
resides exclusively in the prior term. \citet{Cheng_2019_CVPR} provided
a theoretical Bayesian perspective on deep priors, describing them as
Gaussian process priors in classical Bayesian terms. Specifically,
\citet{Cheng_2019_CVPR} showed that for infinitely wide CNNs, i.e.~CNNs
with a large number of channels, the inductive bias of the CNN
architecture and the Gaussian prior on its weights are equivalent to a
stationary Gaussian process prior in the image space.
\citet{Cheng_2019_CVPR} also explicitly made a connection between the
kernel of this Gaussian process and the architecture of a CNN, by
characterizing the effects of convolutions, non-linearities,
up-sampling, down-sampling, and skip connections, which provides
insights on selecting an appropriate CNN architecture. Independently,
\citet{dittmer2020regularization} argue that the weak form of our
constrained formulation with deep priors yield the same solutions for
the correct Lagrange multiplier. This means there is a direct connection
between our formulation and unconstrained variational approaches. The
latter permit a straightforward Bayesian interpretation.

Aside from choosing the right CNN architecture
\citep{Lempitsky, dittmer2020regularization}, random initialization of
its weight \citep{pmlrv9glorot10a} and fixed input for the latent
variable, the above posterior depends on selecting a value for the
tradeoff parameter $\lambda>0$, which weighs the importance of the
Gaussian prior against the noise-variance weighted data misfit term in
the likelihood function. In the sections below, we will comment how to
choose the value for $\lambda$.

The above expression for the posterior in equation~\ref{imaging-obj}
forms the basis of our proposed probabilistic imaging scheme based on
Bayesian inference. Before discussing how to sample from this
distribution, we first briefly describe how to extract various
statistical properties from this posterior distribution on the image.
Specifically, we will review how to obtain point estimates
\citep{casella2021statistical}, including maximum likelihood estimate
(MLE) , maximum a posteriori estimate (MAP) and estimates for the mean
and pointwise standard deviation, and $99\%$ confidence intervals.

\subsection{Estimation with Bayesian
inference}\label{estimation-with-bayesian-inference}

Based on the expressions for the negative log-likelihood
(equation~\ref{deep-prior-likelihood}) and posterior
(equation~\ref{imaging-obj}), we derive expressions for different point
and interval estimates \citep{arridge_oktem_2019}.

\subsubsection{Maximum likelihood
estimation}\label{maximum-likelihood-estimation}

To establish a baseline for image estimates obtained without
regularization, we first consider point estimates for the image that
correspond to finding an image that best fits the observed data. Since
this estimate is obtained by maximizing the likelihood function with
respect to the unknown image, $\delta \B{m}$, this estimate is known as
the MLE. The corresponding optimization problem can be written as
\begin{equation}
\begin{aligned}
\delta \B{m}_{\text{MLE}} & = \mathop{\rm arg\,min}_{\delta \B{m}}
    - \log p_{\text{like}}
    \left ( \B{d} \mid \delta \B{m} \right ) \\
& = \mathop{\rm arg\,min}_{\delta \B{m}}  \frac{1}{2 \sigma^2}
    \sum_{i=1}^{n_s} \big \| \B{d}_i- \B{J}(\B{m}_0, \B{q}_i) \delta \B{m} \big \|_2^2,
\end{aligned}
\label{MLE}
\end{equation}
 where the last equality follows from equation~\ref{imaging-likelihood}.
Observe that the MLE corresponds to the deterministic least-squares
solution, yielded by equation~\ref{imaging-opt}. Unfortunately, MLE
images are prone to overfitting
\citep{casella2021statistical, aster2018parameter} that results in
imaging artifacts \citep{nemeth1999least}.

\subsubsection{Maximum a posteriori
estimation}\label{maximum-a-posteriori-estimation}

Adding regularization to inverse problems, including seismic imaging, is
known to limit overfitting and is capable of filling in at least part of
the null space of the modeling operator. In case of regularization with
deep priors, this corresponds to finding the image that maximizes the
posterior distribution, i.e., we have
\begin{equation}
\begin{aligned}
\B{w}_{\text{MAP}} & = \mathop{\rm arg\,max}_{\B{w}}
    p_{\text{post}} \left ( \B{w} \mid \B{d} \right ) \\
& = \mathop{\rm arg\,min}_{\B{w}} \frac{1}{2 \sigma^2} \sum_{i=1}^{n_s}\big
    \| \B{d}_i- \B{J}(\B{m}_0, \B{q}_i) {g} (\B{z},
    \B{w}) \big \|_2^2 + \frac{\lambda^2}{2}\big  \| \B{w} \big \|_2^2.
\end{aligned}
\label{MAP-w}
\end{equation}
 This estimation for the weights $\B{w}$ is known as the MAP estimate.
Given this estimate $\B{w}_{\text{MAP}}$, the corresponding estimate for
the image is obtained via
\begin{equation}
\delta \B{m}_{\text{MAP}} = g (\B{z}, \B{w}_{\text{MAP}} ).
\label{MAP}
\end{equation}
 When compared with MAP estimates computed from traditional Bayesian
formulations of linear inverse problems, the estimate in
equation~\ref{MAP-w} has several important differences. The above
estimate depends on the random initializations of the weights, $\B{w}$
and latent variable $\B{z}$, which is due to the nonlinearity introduced
by the reparameterization. This renders the above minimization
non-convex, i.e., its local minimum is no longer guaranteed to coincide
with the global minimum. While the objective is non-convex
\citep{NEURIPS2018_a41b3bb3}, as a result of deep prior
reparameterization, because $M\gg N$, first-order stochastic
optimization methods \citep{rmsprop, kingma2014adam, fromage2020} are
able to minimize the objective function in equation~\ref{MAP-w} to small
values of the residual \citep{du2018gradient, pmlr-v97-kunin19a}.
Several other challenges include increased number of iterations,
establishment of a stopping criterion when maximizing
equation~\ref{MAP-w} to prevent overfitting, and the quantification of
uncertainty. Despite these challenge, we argue that invoking the deep
prior outweighs the challenges since it offers a better bias-variance
trade-off and requires knowledge of only a single hyperparameter. In
addition, we refer to \citet{siahkoohi2020SEGwdp} for an alternative
formulation, which reduces the number of iterations and therefore the
number of evaluations of the computationally expensive forward modeling
operator.

\subsubsection{Conditional mean
estimation}\label{conditional-mean-estimation}

So far, the MLE and MAP estimates involved a deterministic (at least for
fixed initialization of the network and latent variable) procedure
maximizing the likelihood or posterior. Since we have access to the
unnormalized posterior PDF,
$p_{\text{post}} \left (\delta \B{w} \mid \B{d} \right)$
(equation~\ref{imaging-obj}), we have in principle ways to retrieve
information on the statistical moments of the posterior distribution of
the unknown perturbation including its mean and pointwise standard
deviation. However, contrary to the two estimates discussed so far these
point estimates can typically only be approximated with samples drawn
from the posterior.

We obtain access to samples from the posterior,
$p_{\text{post}} \left (\delta \B{m} \mid \B{d} \right)$ via a
``push-forward'' of samples from
$p_{\text{post}} \left ( \B{w} \mid \B{d} \right )$ based on the
deterministic map $\delta \B{m} = {g} (\B{z}, \B{w})$ for fixed $\B{z}$
\citep{bogachev2006measure}. As a result, for any sample of the weights,
$\B{w}$, drawn from $p_{\text{post}} \left ( \B{w} \mid \B{d} \right )$,
we have
\begin{equation}
g(\B{z}, \B{w}) \sim p_{\text{post}}
    \left (\delta \B{m} \mid \B{d} \right).
\label{push-forward}
\end{equation}
 Assuming access to $n_w$ samples from the posterior,
$p_{\text{post}} \left ( \B{w} \mid \B{d} \right )$, the first moment,
also known as the conditional mean, can be approximated from these
samples,
$\left \{ \B{w}_j \right \}_{j=1}^{n_{\mathrm{w}}} \sim p_{\text{post}} ( \B{w} \mid \B{d} )$,
via
\begin{equation}
\begin{aligned}
\delta \B{m}_{\text{CM}}
& = \mathbb{E}_{\delta \B{m} \sim p_{\text{post}} \left (\delta \B{m}
    \mid \B{d} \right)} \big [ \delta \B{m} \big ]\\
&= \mathbb{E}_{\B{w} \sim p_{\text{post}}
    \left (\delta \B{w} \mid \B{d} \right)} \big [ g( \B{z}, \B{w}) \big ] \\
& = \int p_{\text{post}} ( \B{w} \mid
    \B{d} ) g( \B{z}, \B{w}) \mathrm{d} \B{w}\\
&\approx \frac{1}{n_{\mathrm{w}}} \sum_{j=1}^{n_{\mathrm{w}}} g(
    \B{z}, \B{w}_j).
\end{aligned}
\label{conditionalmean}
\end{equation}
 We describe the important step of obtaining these samples from the
posterior below.

Compared to the MAP estimate, the conditional mean, which corresponds to
the minimum-variance estimate \citep{anderson1979}, is less prone to
overfitting \citep{mackay2003information}. This was confirmed
empirically for seismic imaging
\citep{siahkoohi2020EAGEdlb, siahkoohi2020SEGuqi}. In the experimental
sections below, we will provide further evidence of advantages the
conditional mean offers compared to MAP estimation.

\subsubsection{Point-wise standard deviation
estimation}\label{point-wise-standard-deviation-estimation}

In its most rudimentary form, uncertainties in the imaging step can be
assessed by computing the pointwise standard deviation, which expresses
the spread among the different unknown models explaining the observed
data. Given samples from the posterior, this quantity can be computed
via
\begin{equation}
\begin{aligned}
\boldsymbol{\sigma}^2_{\text{post}} & = \mathbb{E}_{\delta \B{m}
    \sim p_{\text{post}} \left (\delta \B{m} \mid \B{d} \right) } \big [ (
    \delta \B{m} - \delta \B{m}_{\text{CM}})
    \odot (\delta \B{m} - \delta \B{m}_{
    \text{CM}}) \big ] \\
& \approx \frac{1}{n_{\mathrm{w}}}\sum_{j=1}^{n_{\mathrm{w}}} \big (g(
    \B{z}, \B{w}_j) - \delta
    \B{m}_{\text{CM}} \big ) \odot \big ( g( \B{z},
    \B{w}_j) - \delta \B{m}_{\text{CM}} \big ).
\end{aligned}
\label{pointwise-std}
\end{equation}
 In this expression, $\boldsymbol{\sigma}_{\text{post}}$ is the
estimated pointwise standard deviation and $\odot$ represents
elementwise multiplication. Again, the expectations approximated in
equations~\ref{conditionalmean} and~\ref{pointwise-std} require samples
from the posterior distribution,
$p_{\text{post}} \left (\delta \B{m} \mid \B{d} \right)$.

\subsubsection{Confidence intervals}\label{confidence-intervals}

As described above, the pointwise standard deviation is a quantity that
summarizes the spread among the likely estimates of the unknown. Using
this quantity, we can put error bars on the unknown in which case we
assign probabilities (confidence) to the unknowns being in a certain
interval. The interval is obtained by treating the pointwise posterior
distribution as a Gaussian distribution, where the mean and standard
deviation at each points are equal to the value of the conditional mean
estimate and pointwise standard deviation at that point, respectively.
Given a desired confidence value, e.g., $99\%$, sample mean
$\boldsymbol{\mu}$, and sample variance $\boldsymbol{\sigma}^2$, the
confidence interval is
$\boldsymbol{\mu} \pm 2.576\, \boldsymbol{\sigma}$ where $99\%$ of
samples fall between the left
($\boldsymbol{\mu} - 2.576\, \boldsymbol{\sigma}$) and right
($\boldsymbol{\mu} + 2.576\, \boldsymbol{\sigma}$) tails of the Gaussian
distribution \citep{friedman2001elements}.

\section{Sampling from the posterior
distribution}\label{sampling-from-the-posterior-distribution}

Extracting statistical information from the posterior distribution, such
as the point and interval estimates introduced in the previous section,
typically requires access to samples from the posterior distribution. In
the following section, we first show that approximations to the point
and interval estimates are instances of Monte Carlo integration, given
samples from the posterior distribution. Next, we shift our attention to
constructive techniques to draw these samples efficiently by introducing
preconditioning and crucial strategies to select the stepsize. Finally,
we describe an empirical verification of convergence of the Markov
chains that we will use to verify our sampling approach.

\subsection{Monte Carlo sampling}\label{monte-carlo-sampling}

For most applications the posterior PDF is not directly of interest, but
we need to evaluate expectations involving the posterior distribution
instead. Given samples from the
posterior,$\left \{ \B{w}_j \right \}_{j=1}^{n_{\mathrm{w}}} \sim p_{\text{post}} ( \B{w} \mid \B{d} )$,
these expectations with respect to arbitrary functions can be
approximated by
\begin{equation}
\mathbb{E}_{\B{w} \sim p_{\text{post}} \left (\delta \B{w} \mid \B{d} \right)}
    \big [ f (\B{w} ) \big ] \approx \frac{1}{n_w}
    \sum_{j=1}^{n_w} f (\B{w}_{j} ).
\label{monte-carlo}
\end{equation}
 Below we describe our proposed MCMC approach for obtaining samples from
the posterior.

\subsection{Sampling via stochastic gradient Langevin
dynamics}\label{sampling-via-stochastic-gradient-langevin-dynamics}

Drawing samples from posterior distributions associated with imaging
problems of high dimensionality ($M,\, N$ large) and expensive forward
operators (e.g., demigration operators) is challenging
\citep{welling2011bayesian, MartinMcMC2012}. Among the different
approaches, MCMC is a well-studied technique capable of drawing samples
via a sequential random-walk procedure. This process requires evaluation
of the posterior PDF at each step. The need for repeated evaluations of
the forward operator, the correlation between consecutive samples
\citep{gelman2013bayesian}, and the high dimensionality of the problem
are the chief computational challenges for these methods. Despite these
difficulties, MCMC methods have been applied successfully in imaging
problems
\citep{curtis2001prior, fang2018uqfip, herrmann2019NIPSliwcuc, zhao2019gradient, kotsi2020, siahkoohi2020EAGEdlb, siahkoohi2020SEGuqi}.

Aside from problems related to the required length of the Markov Chains,
computing the misfit over all $n_s$ sources in the likelihood term of
the posterior PDF (equation~\ref{imaging-obj}) is problematic since this
calls for many evaluations of the linearized Born scattering operator.
To address this issue, we use techniques from stochastic optimization
\citep{robbins1951stochastic, nemirovski2009robust, li2018plane} where
the gradients are evaluated for a single randomly selected source
(without replacement) at each iteration. For first-order methods, this
technique is known as stochastic gradient descent
\citep[SGD;][]{robbins1951stochastic} and widely used in the machine
learning and wave-based inversion communities
\citep{vanLeeuwen2010IJGswi, haber2012effective, herrmann2012, rmsprop, kingma2014adam, lu2015, tu2015, li2018}.

While SGD bring down the computational costs, it is a stochastic
optimization algorithm for finding the mode of the posterior
distribution and it does not provide samples from the posterior
distribution. In order to do that, we have to add a carefully calibrated
noise term to the gradients. This additional noise term induces a random
walk from which samples from the posterior distribution can be drawn
under certain conditions \citep{welling2011bayesian}. Adding this noise
term also avoids converge of the iterations to the MAP estimate
\citep{welling2011bayesian}. In this paper, we adapt an approach known
as stochastic gradient Langevin dynamics
\citep[SGLD,][]{welling2011bayesian}, which is designed to reduce the
number of necessary individual likelihood evaluations at each iteration.
SGLD was originally developed for Bayesian inference on deep neural
networks trained on large-scale datasets. Compared to the original
formulation of Langevin dynamics \citep{neal2011mcmc}, SGLD works on
randomly selected subsets of shot data, which makes it computationally
more efficient and achievable at least in 2D imaging problems.
Asymptotically, SGLD provides accurate samples from the target
distribution
\citep{welling2011bayesian, pmlr-v32-satoa14, raginsky2017non, NIPS2018_8048, NEURIPS2020_b5b8c484}---in
our case, the imaging posterior distribution. It differs from
variational inference \citep{jordan1999introduction} in that no
surrogate distribution is formulated and matched to the distribution of
interest.

Following the work of \citet{welling2011bayesian}, SGLD iterations for
the negative log-posterior involve at iteration $k$ the following update
for the network weights of the deep prior:
\begin{equation}
\begin{aligned}
& \B{w}_{k+1} = \B{w}_{k} - \frac{\alpha_k}{2} \B{M}_k
    \nabla_{\B{w}} \left ( \frac{n_s }{2 \sigma^2} \big \| \B{d}_i-
    \B{J}(\B{m}_0, \B{q}_i) {g} (\B{z},
    \B{w}_k) \big \|_2^2  + \frac{\lambda^2}{2} \big \| \B{w}_k \big \|_2^2
    \right ) + \boldsymbol{\eta}_k, \\
& \boldsymbol{\eta}_k \sim \mathrm{N}( \B{0}, \alpha_k \B{M}_k),
\end{aligned}
\label{sgld}
\end{equation}
 where the index, $i\subset \{1, \ldots, n_s\}$, is chosen randomly
without replacement at each iteration. Once all the shots are drawn, we
start all over by redrawing indices, without replacement, from
$i\subset \{1, \ldots, n_s\}$. We repeat this process for $K$ steps (see
Algorithm 1), where $K$ can be arbitrarily large. To ensure and speedup
convergence, the stepsizes $\alpha_k$ and the adaptive preconditioning
matrix $\B{M}_k$ need to be chosen carefully. The additional zero-mean
Gaussian noise term $\boldsymbol{\eta}_k $ with covariance matrix
$\alpha_k \B{M}_k$ distinguishes between the update rule in
equation~\ref{sgld} and SGD optimization algorithm. It was shown by
\citep{welling2011bayesian} that the above iterations sample from the
posterior after a warmup phase, i.e., a certain number of iterations of
equation~\ref{sgld}. During the warmup stage, these iterations behave
similarly to those of the SGD algorithm but at some point transition to
the proper sampling phase \citep{welling2011bayesian}. Below we will
comment when that transition is likely to occur.

\subsubsection{Stepsize selection}\label{stepsize-selection}

Convergence of stochastic optimization methods such as SGD and SGLD
relies on carefully designed stepsize strategies. Compared to SGD, SGLD
has the additional complication of having to balance random errors due
to randomly selecting shot records and the deliberate random ``errors''
induced by the additional Gaussian noise term, $\boldsymbol{\eta}_k$. On
the one hand, the iterations in equation~\ref{sgld} need to make
sufficient progress during the warmup phase so that the samples ($=$
iterations $\B{w}_k$) become independent of the chain's initialization,
i.e., the weights $\B{w}_0$ at the start. On the other hand, after
warmup the Gaussian noise term, $\boldsymbol{\eta}_k$ will start to
dominate the energy of the error in the gradient caused by the
stochastic approximation to the likelihood function
(equation~\ref{deep-prior-likelihood}). This can be explained by the
fact that the variance of error due to the stochastic gradient
approximation is proportional to the square of the stepsize
\citep{robbins1951stochastic}, whereas the additive noise term is drawn
from a Gaussian distribution whose variance is proportional to the
stepsize. Consequently, for small stepsizes, it is expected that the
error in gradients will be dominated by additive noise
\citep{welling2011bayesian}, which effectively turns equation~\ref{sgld}
to Langevin dynamics \citep{neal2011mcmc}. As a result, similar to SGD,
convergence can only be guaranteed when the stepsize in
equation~\ref{sgld} decreases to zero. However, this would increase the
number of iterations to fully explore the posterior probability space.
We avoid this situation and follow \citet{welling2011bayesian} who
propose the following sequence of stepsizes:
\begin{equation}
\alpha_{k}= a (b+k)^{-\gamma},
\label{stepsize}
\end{equation}
 where $\gamma = \frac{1}{3}$ is the decay rate chosen according to
\citet{teh2016consistency}. The constants $a,\ b$ in this expression
control the initial and final value of the stepsize. Below, we will
comment how to chose these constants and how to ensure that potential
posterior sampling errors \citep{NIPS2018_8048} are avoided.

\subsubsection{Preconditioning}\label{preconditioning}

In addition to selecting proper stepsizes, the converge of the
iterations in equation~\ref{sgld} depends on how strongly the different
weights of the deep prior are coupled to the data. Without
preconditioning, i.e., $\B{M}_k = \B{I}$, SGLD updates all parameters
with one and the same stepsize. This leads to slow convergence of the
insensitive weights that are weakly coupled to the data. To avoid this
situation, \citet{li2016preconditioned} proposed an adaptive diagonal
preconditioning matrix extending the RMSprop optimization algorithm
\citep{rmsprop}. This preconditioner is deigned to speed up the initial
warmup and subsequent sampling stage of the iterations in
equation~\ref{sgld}. To define this preconditioning matrix, let
$\delta \B{w}$ denote the gradient of the negative log-posterior density
(equation~\ref{imaging-obj}) at the current estimate of weights
$\B{w}_k$, i.e.,
\begin{equation}
\delta\B{w} = \nabla_{\B{w}} \left ( \frac{n_s }{2 \sigma^2}
    \big \| \B{d}_i- \B{J}(\B{m}_0, \B{q}_i) {g} (\B{z}, \B{w}_k) \big \|_2^2
    + \frac{\lambda^2}{2} \big \| \B{w}_k \big \|_2^2 \right ).
\label{gradient}
\end{equation}
 Given these gradients, define the following running pointwise sum on
the pointwise square of the gradients
\begin{equation}
\B{v}_{k+1} =\beta \B{v}_{k} + \left (1 - \beta \right ) \delta\B{w} \odot \delta\B{w}
\label{ruunning-grad}
\end{equation}
 where the parameter $\beta$ controls the relative importance of the
elementwise square of the gradient compared to the current iterate
$\B{v}_{k}$. The $\B{v}_0$ is initialized as a vector with $M$ zeros. By
choosing,
\begin{equation}
\B{M}_k = \text{diag} \left ( 1 \oslash \sqrt{\B{v}_{k+1}} \right )
\label{precond-mat}
\end{equation}
 with $\oslash$ elementwise division, the effective stepsize for network
weights with large (on average) gradients, i.e., large sensitivities, is
lowered whereas weights with small (on average) gradients get updated
with a larger effective stepsize. To avoid division by zero, we add a
small value to the denominator of equation~\ref{precond-mat}. By
introducing the preconditioning matrix $\B{M}_k$ all weights are updated
similarly, which allows us to increase the stepsize. Following
\citet{li2016preconditioned}, we set $\beta = 0.99$. In addition to
leveling the playing field, for the gradients themselves the
preconditioning matrix also scales the essential additive noise term so
the random walk proceeds isotropically.

\subsection{Practical verification}\label{practical-verification}

While there exists a well established literature on how to verify
whether Markov chains produce accurate samples from the posterior
distribution \citep[see][]{gelman_rubin_92}, these methods are typically
impractical for our problem. We will adopt a more pragmatic approach to
assess the accuracy of the samples drawn from our Markov chains computed
with SGLD, as it will be explained in the following.

We first validate the accuracy of sampling from a single Markov chain by
computing confidence intervals. These intervals are computed from
posterior samples obtained via one MCMC chain using SGLD
(equation~\ref{sgld}). By definition, these confidence intervals provide
the range within which the weights and therefore the image are expected
to fall. This means that MAP estimates for the seismic image should
ideally fall within these confidence intervals computed from the
posterior samples. While the variability among MAP estimates is less
than the variability among true posterior samples, we still find this
test of practical importance. To verify this, we compute multiple MAP
estimates (see equations~\ref{MAP-w} and~\ref{MAP}) for different
independent random initializations of the deep prior weights, $\B{w}$.
MAP estimates are obtained via stochastic optimization using the RMSprop
optimization algorithm \citep{rmsprop}, which uses the same
preconditioning scheme (equations~\ref{ruunning-grad}
and~\ref{precond-mat}) as SGLD. By checking whether the different MAP
estimates indeed fall within the computed confidence interval, the
accuracy of the samples from the posterior can at least be verified
qualitatively.

Ideally, different Markov chains initialized with different weights
should lead to similar statistics for samples of the posterior
distribution. We verify this empirically by running chains with
different independently randomly initialized weights, followed by visual
inspection of the conditional mean and pointwise standard deviation
derived from samples generated by the different chains. Deviations among
the estimates provides us with at least an indication of areas in the
image where we should be less confident on the inferred statistics.

\subsection{The SGLD Algorithm}\label{the-sgld-algorithm}

The different steps of generating $n_w=K/2$ samples from the posterior
from $K$ iterations of SGLD (equation~\ref{sgld}) are summarized in
Algorithm~\ref{alg}~for a given set of $n_s$ processed shot records and
their respective source signatures,
$\left \{ \B{d}_{i}, \B{q}_{i} \right \}_{i=1}^{n_s}$. Aside from shot
data, Algorithm~\ref{alg}~requires a smooth background model, $\B{m}_0$,
for the squared slowness and a fixed realization for the latent variable
$\B{z} \sim \mathrm{N}( \B{0}, \B{I})$. In addition to these input
vectors, SGLD requires hyperparameters to be set for the

\begin{itemize}
\item
  \textbf{Stepsize strategy.} Following \citet{teh2016consistency}, the
  decay rate parameter in equation~\ref{stepsize} is set to
  $\gamma=\frac{1}{3}$. The stepsize constants $a,\ b$ in
  equation~\ref{stepsize} are chosen separately for each presented
  numerical experiment to ensure fast convergence in the warmup phase.
  Specifically, we select $a$ large enough to ensure fast convergence
  while making sure the initial iterations do not diverge due to large a
  stepsize. We selected $b$ to be the same as the stepsize that would
  yeld a good convergence for the MAP estimation problem, i.e., SGLD
  iterations without the additive noise. This is to ensure SGLD
  iterations get close enough to mode(s) of the distribution toward the
  end. While selecting these parameters differently changes the speed of
  converge, the accuracy of the resulting samples is empirically
  verified for the chosen parameters.
\item
  \textbf{Preconditioning.} As documented in the literature, we chose
  $\beta=0.99$ in equation~\ref{ruunning-grad}.
\item
  \textbf{Noise variance.} The variance $\sigma^2$ of the noise assumed
  to be known.
\item
  \textbf{Regularization parameter.} As with many inverse problems, the
  selection of the regularization parameter $\lambda^{-2}$ is
  challenging. While sophisticated techniques \citep{aster2018parameter}
  exist to estimate this parameter, we tune the regularization parameter
  $\lambda^{-2}$ by hand to limit the imaging artifacts visually.
\item
  \textbf{Number of iterations and warmup.} We run $10\,$k SGLD
  iterations (equation~\ref{sgld}) in total, and we adopt the general
  practice of discarding the first half of the obtained samples
  \citep{gelman_rubin_92}.
\end{itemize}

Given the above inputs, Algorithm~\ref{alg}~proceeds by running $K$
iterations during which simultaneous shot records, each made of a
Gaussian weighted source aggregate, are selected, followed by
calculations of the gradient (line $3$), calculation of the
preconditioner (lines $4-5$), stepsize (line $6$), and update of the
weights (line $8$). After $K/2$ iterations, the updated weight also
serve as samples from the posterior
\citep{gelman_rubin_92, welling2011bayesian}.

\begin{scholmdAlgorithm}
\textbf{Input:}~\vspace{.01em}\\\hspace*{0.333em}\hspace*{0.333em}\hspace*{0.333em}$\left \{ \B{d}_{i}, \B{q}_{i} \right \}_{i=1}^{n_s}$~~~~\hspace*{\fill}~\texttt{//\phantom{\ }observed\phantom{\ }data\phantom{\ }and\phantom{\ }source\phantom{\ }signatures}\vspace{.01em}\\\hspace*{0.333em}\hspace*{0.333em}\hspace*{0.333em}$\B{m}_0$~~~~\hspace*{\fill}~\texttt{//\phantom{\ }smooth\phantom{\ }background\phantom{\ }squared-slowness\phantom{\ }model}\vspace{.01em}\\\hspace*{0.333em}\hspace*{0.333em}\hspace*{0.333em}$\B{z} \sim \mathrm{N}( \B{0}, \B{I})$~~~~\hspace*{\fill}~\texttt{//\phantom{\ }fixed\phantom{\ }input\phantom{\ }to\phantom{\ }the\phantom{\ }CNN}\vspace{.01em}\\\hspace*{0.333em}\hspace*{0.333em}\hspace*{0.333em}$\lambda^{-2}$~~~~\hspace*{\fill}~\texttt{//\phantom{\ }variance\phantom{\ }of\phantom{\ }Gaussian\phantom{\ }prior\phantom{\ }on\phantom{\ }CNN\phantom{\ }weights}~\vspace{.01em}\\\hspace*{0.333em}\hspace*{0.333em}\hspace*{0.333em}$\sigma^2$~~~~\hspace*{\fill}~\texttt{//\phantom{\ }estimated\phantom{\ }noise\phantom{\ }variance}~\vspace{.01em}\\\hspace*{0.333em}\hspace*{0.333em}\hspace*{0.333em}$\beta$~~~~\hspace*{\fill}~\texttt{//\phantom{\ }weighting\phantom{\ }parameter\phantom{\ }for\phantom{\ }constructing\phantom{\ }the\phantom{\ }preconditioning\phantom{\ }matrix}~\vspace{.01em}\\\hspace*{0.333em}\hspace*{0.333em}\hspace*{0.333em}$a, b$~~~~\hspace*{\fill}~\texttt{//\phantom{\ }stepsize\phantom{\ }parameters\phantom{\ }in\phantom{\ }equation}~\ref{stepsize}~\vspace{.01em}\\\hspace*{0.333em}\hspace*{0.333em}\hspace*{0.333em}$K$~~~~~\hspace*{\fill}~\texttt{//\phantom{\ }maximum\phantom{\ }MCMC\phantom{\ }steps}~\vspace{.6em}\\\textbf{Initialization:}~\vspace{.01em}\\\hspace*{0.333em}\hspace*{0.333em}\hspace*{0.333em}randomly~initialize~CNN~parameters,~$\B{w}_{0}\in\R^M$~~~~~\hspace*{\fill}~\citet{pmlrv9glorot10a}~\vspace{.01em}\\\hspace*{0.333em}\hspace*{0.333em}\hspace*{0.333em}initialize~vector,~$\B{v}_0\in\R^M$~with~zero~~\vspace{.6em}\\1.~\textbf{for}~$k=0$~\textbf{to}~$K-1$~\textbf{do}~\vspace{.4em}\\2.~~~~randomly~draw~$i\subset \{1, \ldots, n_s\}$~~~~\hspace*{\fill}~\texttt{//\phantom{\ }sample\phantom{\ }without\phantom{\ }replacement}~\vspace{.1em}\\3.~~~~$\delta\B{w} = \nabla_{\B{w}} \left ( \frac{n_s }{2 \sigma^2} \big \| \B{d}_i- \B{J}(\B{m}_0, \B{q}_i) {g} (\B{z}, \B{w}_k) \big \|_2^2 + \frac{\lambda^2}{2} \big \| \B{w}_k \big \|_2^2 \right )$~~~~\hspace*{\fill}~\texttt{//\phantom{\ }equation}~\ref{gradient}~\vspace{.1em}\\4.~~~~$\B{v}_{k+1} =\beta \B{v}_{k} + \left (1 - \beta \right ) \delta\B{w} \odot \delta\B{w}$~~~~\hspace*{\fill}~\texttt{//\phantom{\ }equation}~\ref{ruunning-grad}~\vspace{.3em}\\5.~~~~$\B{M}_k = \text{diag} \left ( 1 \oslash \sqrt{\B{v}_{k+1}} \right )$~~~~\hspace*{\fill}~\texttt{//\phantom{\ }equation}~\ref{precond-mat}~\vspace{.3em}\\6.~~~~$\alpha_{k}= a (b+k)^{-\gamma}$~~~~\hspace*{\fill}~\texttt{//\phantom{\ }equation}~\ref{stepsize}~\vspace{.3em}\\7.~~~~$\boldsymbol{\eta}_k \sim \mathrm{N}( \B{0}, \alpha_k \B{M}_k)$~~~~\hspace*{\fill}~\texttt{//\phantom{\ }draw\phantom{\ }noise\phantom{\ }to\phantom{\ }add\phantom{\ }to\phantom{\ }gradient}~\vspace{.3em}\\8.~~~~$\B{w}_{k+1} = \B{w}_{k} - \frac{\alpha_k}{2} \B{M}_k \delta\B{w} + \boldsymbol{\eta}_k$~~~~\hspace*{\fill}~\texttt{//\phantom{\ }update\phantom{\ }rule\phantom{\ }according\phantom{\ }to\phantom{\ }equation}~\ref{sgld}~\vspace{.3em}\\9.~\textbf{end~for}~\vspace{.6em}\\\textbf{Output:}~$\left \{ \B{w}_{k} \right \}_{k=K/2+1}^K$~~~~\hspace*{\fill}~\texttt{//\phantom{\ }samples\phantom{\ }from\phantom{\ }the\phantom{\ }posterior}~$p_{\text{post}} ( \B{w} \mid \B{d} )$~\vspace{.3em}
\caption{Seismic imaging posterior sampling with SGLD.}\label{alg}
\end{scholmdAlgorithm}

\section{Validating Bayesian
inference}\label{validating-bayesian-inference}

In this section, we validate our approach with synthetic examples. To
mimic a realistic imaging scenario where the ground truth is known, we
do this on a ``quasi''-field data set, made out of noisy synthetic shot
data generated from a real migrated image. After demonstrating the
benefits of regularization with the deep prior, we compare the MAP
estimate with the conditional mean. The latter minimizes the Bayesian
risk, i.e., it minimizes in expectation the $\ell_2$-norm squared
difference between the true image and inverted image, given shot data
\citep{anderson1979}. We conclude by reviewing the pointwise standard
deviation as a measure of uncertainty, which can be reaped from samples
drawn from the posterior.

\subsection{Problem setup}\label{problem-setup}

\begin{figure*}
\centering
\subfloat[\label{true-model}]{\includegraphics[width=0.950\hsize]{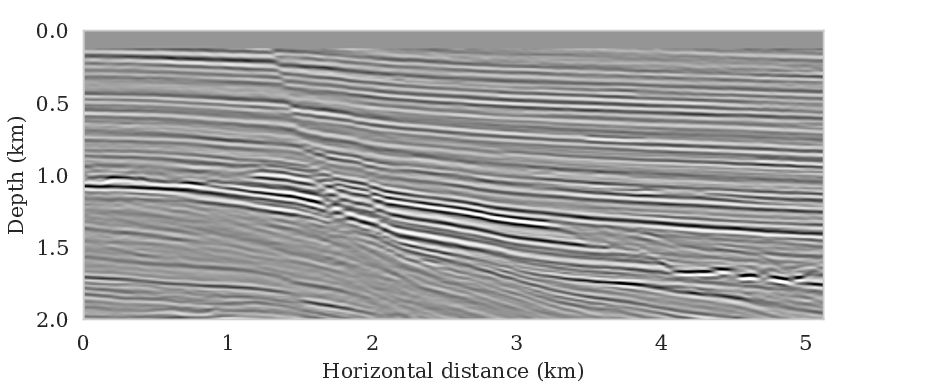}}
\\
\subfloat[\label{smooth-model}]{\includegraphics[width=0.950\hsize]{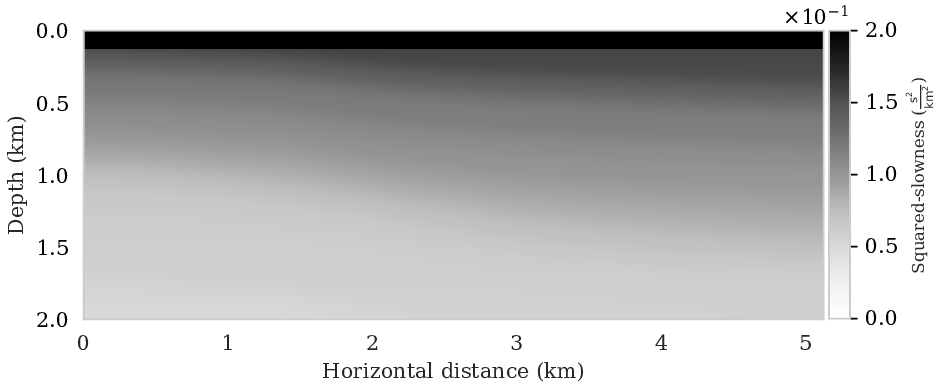}}
\caption{Problem setup. (a) A 2D subset of the Parihaka dataset,
considered as true model. (b) Made up smooth squared-slowness background
model.}\label{problem-steup}
\end{figure*}

With few exceptions, synthetic models often miss realistic statistics of
the spatial distribution of the seismic reflectivity. To avoid working
with over simplified seismic images, we generate ``quasi''-field shot
data derived from a 2D subset of the real prestack Kirchhoff time
migrated \href{https://wiki.seg.org/wiki/Parihaka-3D}{Parihaka-3D}
dataset \citep{Veritas2005, WesternGeco2012} released by the New Zealand
government. We call our experiment ``quasi'' real because synthetic data
is generated from migrated field data that serves as a proxy for the
unknown true medium perturbations (Figure~\ref{true-model}). Due to the
nature of the migration algorithm used to obtain the Parihaka dataset,
the amplitudes in the extracted 2D subset are not necessarily consistent
with the seismic imaging forward model presented in this paper
(equation~\ref{linear-fwd-op}). For this reason, we normalized the
amplitudes of the extracted seismic image. Given these perturbations,
shot data is generated with the linearized Born scattering operator for
a made up, but realistic, smoothly varying background model $\B{m}_0$
for the squared slowness (Figure~\ref{smooth-model}). To ensure good
coverage, $205$ shot records are simulated and sampled with a source
spacing of $25\, \mathrm{m}$. Each shot is recorded over $1.5$ seconds
with $410$ fixed receivers sampled at $12.5\, \mathrm{m}$ spread across
full survey area. The source is a Ricker wavelet with a central
frequency of $30\, \mathrm{Hz}$.

\begin{figure*}
\centering
\subfloat[\label{d-noise-free}]{\includegraphics[width=0.350\hsize]{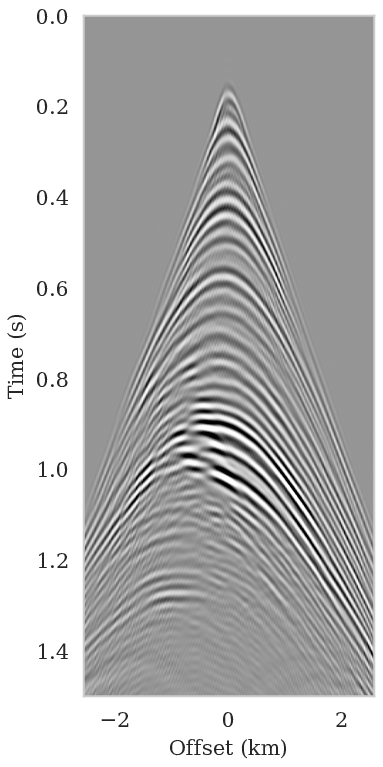}}
\subfloat[\label{d-colored-noise}]{\includegraphics[width=0.350\hsize]{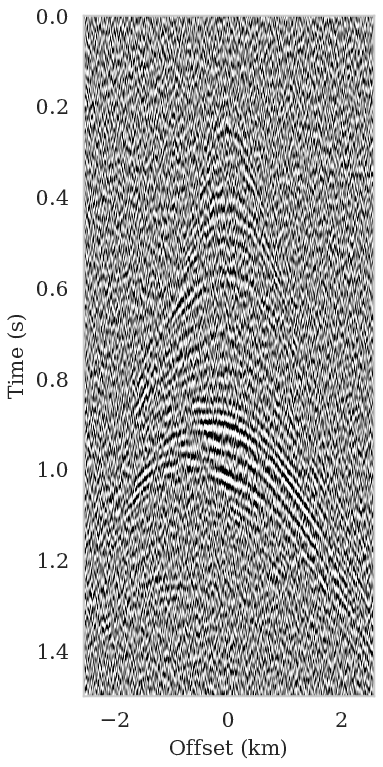}}
\caption{A shot record generated from an image extracted from the
Parihaka dataset. (a) Noise-free linearized data. (b) Linearized data
with band-limited noise.}\label{data}
\end{figure*}

We also add a significant amount of band-limited noise to the shot data
by filtering Gaussian white noise with the source wavelet. The resulting
signal-to-noise ratio for all data is $-8.74\, \mathrm{dB}$, which is
low. Figure~\ref{data} shows an example of a single noise-free
(Figure~\ref{d-noise-free}) and noisy (Figure~\ref{d-colored-noise})
shot record.

Even though our example is in 2D, the number of parameters (the weights
of the deep prior network) is large (approximately $40$ times larger
than image dimension), which results in many SGLD iterations. In a
setting where we are content with approximate Bayesian inference, i.e.,
where the validity of the Markov chains can be established qualitatively
in the way described earlier, we found that ten thousand iterations are
adequate. We adopt the general practice of discarding the first half the
MCMC iterations (about $25$ passes over the data)
\citep{gelman2013bayesian}, which leaves five thousand iterations
dedicated to posterior sampling phase \citep{zhu2018seismic}. The
stepsize sequence is chosen according to equation~\ref{stepsize} with
$a,\ b$ chosen such the stepsize decreases from $10^{-2}$ to
$5 \times 10^{-3}$.

\subsection{Imaging with versus without the deep
prior}\label{imaging-with-versus-without-the-deep-prior}

For reference, we first compare imaging results with and without
regularization. The latter is based on maximizing the likelihood
(equation~\ref{MLE}) whereas the former involves maximizing the
posterior distribution (equation~\ref{MAP-w}). To prevent overfitting of
the MLE estimate, the number of iterations is limited to the equivalent
of only four data passes (four loops over all shots). To ensure
convergence, the number of data passes (or epochs) for the MAP estimate
was set to $15$ (about three thousand iterations). Since the ground
truth is known, the optimal value for the
$\lambda^{-2}=5 \times 10^{-3}$ was found by grid search and picking the
value that visually limits imaging artifacts. Results of minimizing the
negative log-likelihood and negative log-posterior are included in
Figures~\ref{mle-colored} and~\ref{map-colored}, respectively. We
obtained these results with the RMSprop optimization algorithm
\citep{rmsprop}, which uses the same preconditioning scheme
(equations~\ref{ruunning-grad} and~\ref{precond-mat}) as SGLD that we
will use later to conduct Bayesian inference. Compared to vanilla SGD
with a fixed stepsize, RMSprop is an adaptive stepsize method conducive
to the preconditioner introduced in equations~\ref{ruunning-grad}
and~\ref{precond-mat}. As with the SGLD updates, the gradient
calculations involve a single randomly selected shot record. As
expected, compared to the MAP estimate with signal-to-noise ratio (SNR)
$8.79\,$dB, the MLE estimate (SNR $8.25\,$dB) lacks important details,
e.g., weak reflectors in deeper sections, and exhibits strong artifacts,
including imaged reflectors that are noisy and lack continuity. The
latter is important since the estimated seismic image will be used to
automatically track horizons.

\subsection{Bayesian inference with deep
priors}\label{bayesian-inference-with-deep-priors}

As the comparison between MLE and MAP estimates clearly showed,
regularization improves the image but important issues remain. First,
the use of deep priors can lead to overfitting even when a Gaussian
prior on the weights is included. As reported in the literature
\citep{Lempitsky, Cheng_2019_CVPR}, stopping early can be a remedy but a
stopping criterion remains elusive rendering this type of regularization
less effective. Second, the uncertainty is not captured by the MAP
estimation. As we will demonstrate, the ability to draw samples from the
posterior remedies these issues.

\subsubsection{Conditional mean}\label{conditional-mean}

\begin{figure*}
\centering
\subfloat[\label{mle-colored}]{\includegraphics[width=0.900\hsize]{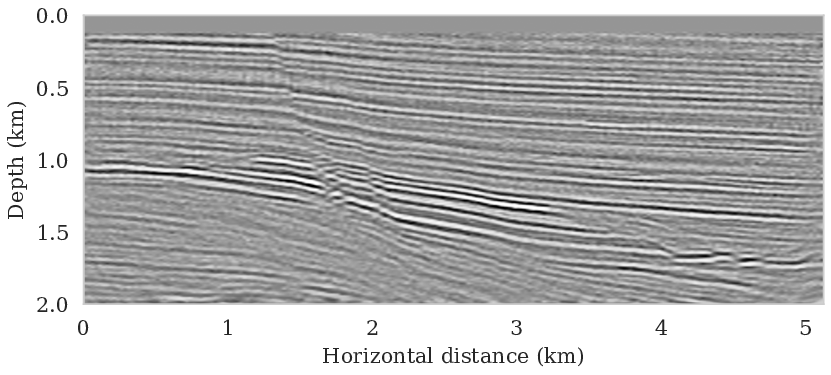}}
\\
\subfloat[\label{map-colored}]{\includegraphics[width=0.900\hsize]{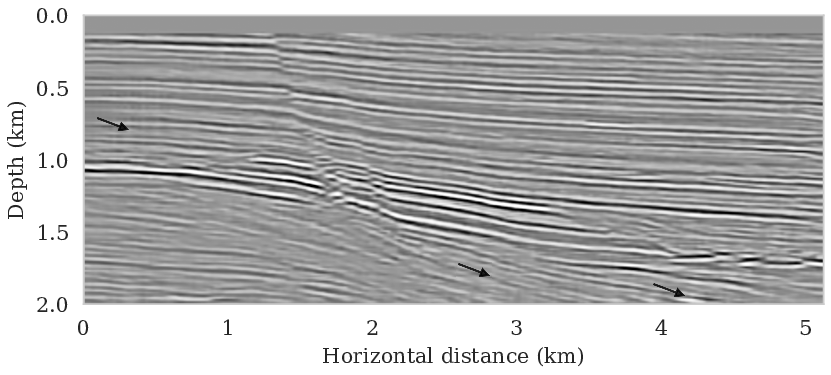}}
\\
\subfloat[\label{cm-colored}]{\includegraphics[width=0.900\hsize]{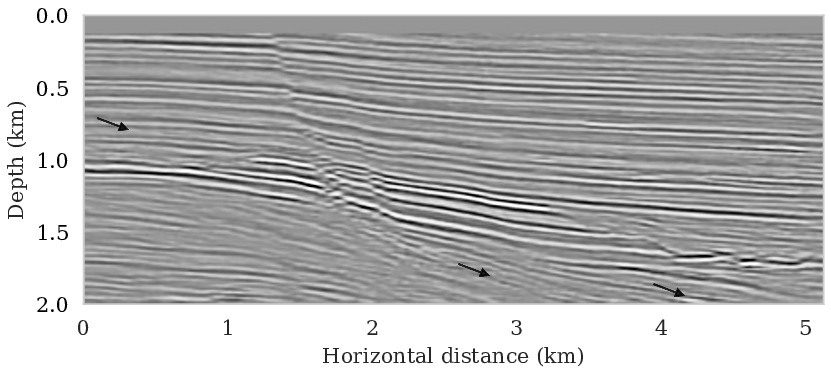}}
\caption{Imaging with deep priors of a 2D subset of the Parihaka dataset
(a) MLE, i.e., minimizer of equation~\ref{imaging-likelihood} with
respect to $\delta \B{m}$, with SNR $8.25\,$dB. (b) The MAP estimate,
i.e., minimizer of equation~\ref{imaging-obj}, followed by a mapping
onto the image space via $g$ (equation~\ref{MAP}), with SNR $8.79\,$dB.
(c) The conditional (posterior) mean estimate,
$\delta \B{m}_{\text{CM}}$, with SNR $9.66\,$dB. All figures are
displayed with the same color clipping values.}\label{mle-vs-map-vs-cm}
\end{figure*}

As described earlier, samples from the posterior provide access to
useful statistical information including approximations to moments of
the distribution such as the mean. With the minor modifications proposed
by \citet{li2016preconditioned} to the RMSprop optimization algorithm,
the posterior distribution can be sampled with Algorithm~\ref{alg} after
a warmup phase of about $25$ data passes. The resulting samples for the
weights are then used, after push forward (see
equation~\ref{push-forward}), to approximate the conditional mean,
$\delta \B{m}_{\text{CM}}$, by computing the sum in
equation~\ref{conditionalmean}. Compared to the MAP estimate
(Figures~\ref{map-colored} and~\ref{cm-colored}), the
$\delta \B{m}_{\text{CM}}$ (SNR $9.66\,$dB) is tantamount to another
significant improvement especially for weaker reflectors in the deeper
part of the image and for reflectors denoted by the arrows.

While there has been a debate in the literature on the accuracy of the
MAP versus conditional mean estimates in the context of regularization
with handcrafted priors, such as total variation
\citep{burger2014maximum}, we find that the conditional mean estimate
negates the need to stop early and is also more robust with respect to
noise.

\subsubsection{Pointwise standard deviation and
histograms}\label{pointwise-standard-deviation-and-histograms}

\begin{figure*}
\centering
\subfloat[\label{std-colored}]{\includegraphics[width=0.900\hsize]{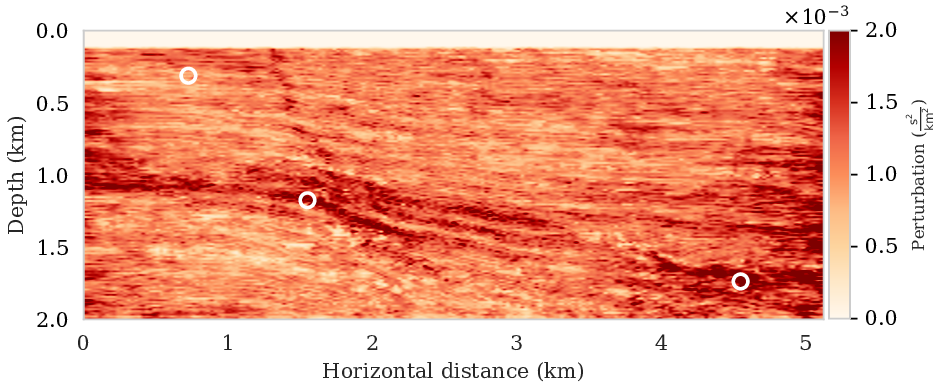}}
\\
\subfloat[\label{std-normalized}]{\includegraphics[width=0.900\hsize]{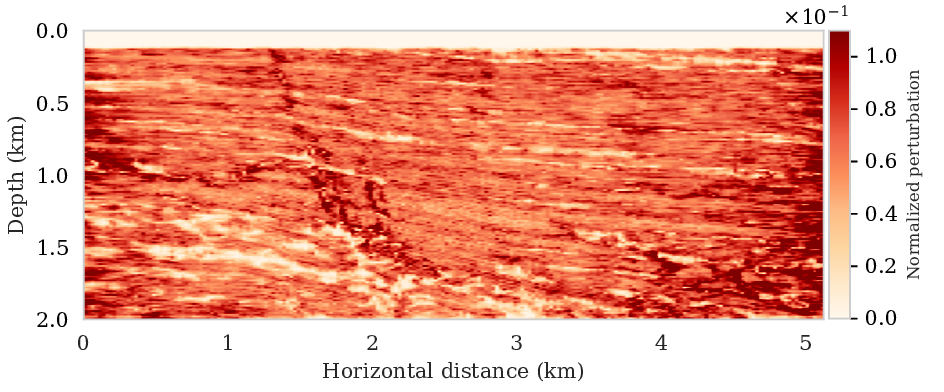}}
\caption{Imaging uncertainty quantification for the Parihaka example.
(a) The pointwise standard deviation among samples drawn from the
posterior, $\boldsymbol{\sigma}_{\text{post}}$. (b) Normalized pointwise
standard deviation by the conditional mean estimate
(Figure~\ref{cm-colored}).}\label{std-plots}
\end{figure*}

To assess variability among the different samples from the posterior, we
include a plot of the pointwise standard deviation
$\boldsymbol{\sigma}_{\text{post}}$ (equation~\ref{pointwise-std}) in
Figure~\ref{std-colored}. This quantity is a measure for uncertainty. To
avoid bias by strong amplitudes in the estimated image, we also plot the
stabilized division of the standard deviation by the envelope of the
conditional mean in Figure~\ref{std-normalized}. From these plots in
Figure~\ref{std-plots}, we observe that as expected uncertainty is large
in areas with a complex geology, e.g., along the faults and along the
tortuous reflectors, and in areas with relative poor illumination deep
in the image and near the edges. On the other hand, the shallow areas of
the image exhibit low uncertainty, which is to be expected due to
proximity to the sources and receivers.

\begin{figure*}
\centering
\subfloat[\label{hist-29x25-colored}]{\includegraphics[width=0.850\hsize]{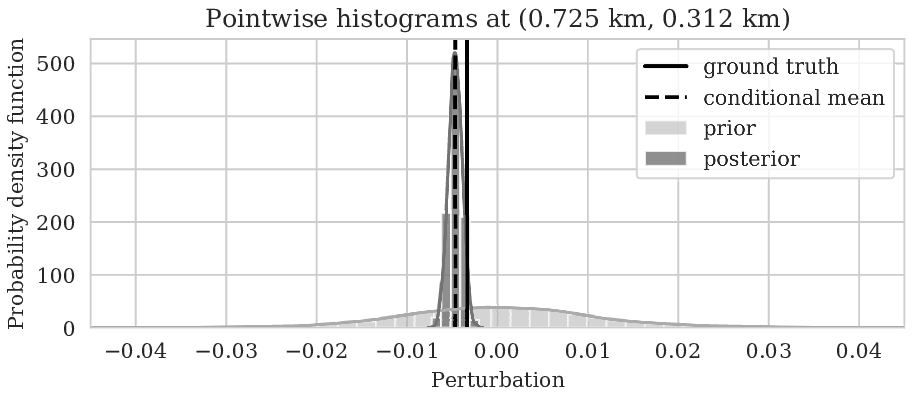}}
\\
\subfloat[\label{hist-62x94-colored}]{\includegraphics[width=0.850\hsize]{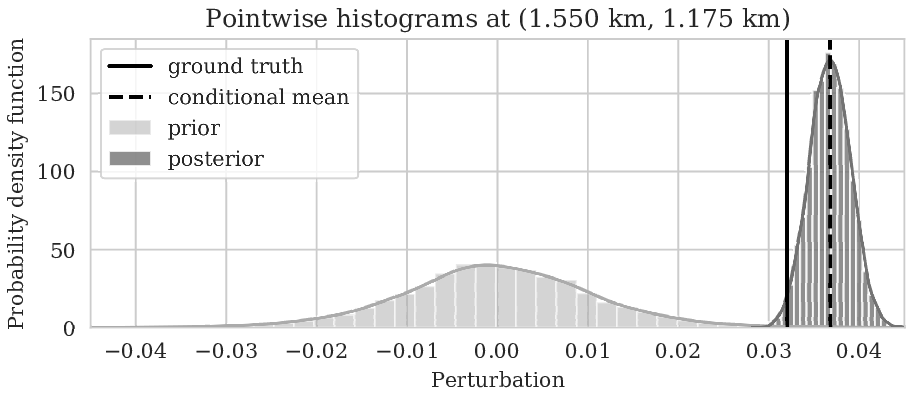}}
\\
\subfloat[\label{hist-182x139-colored}]{\includegraphics[width=0.850\hsize]{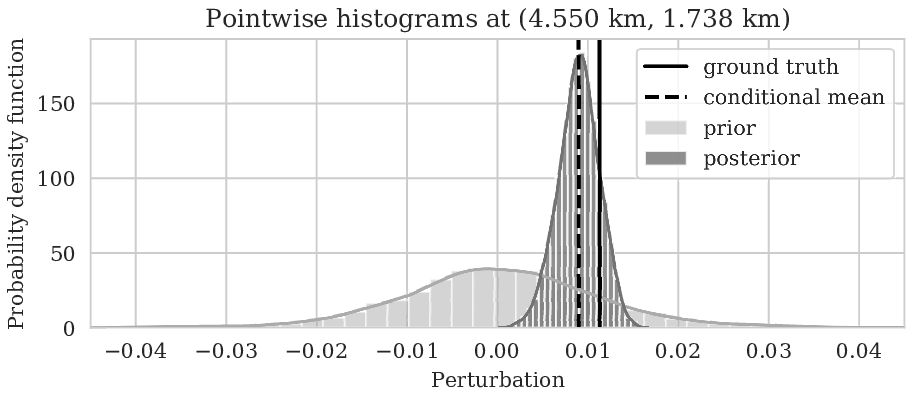}}
\caption{Pointwise prior (light gray) and posterior (dark gray)
histograms along with the true perturbation values (solid black line)
and conditional mean (dashed black line) for points located at (a)
$(0.725\, \mathrm{km}, \ 0.312\, \mathrm{km})$, (b)
$(1.550\, \mathrm{km}, \ 1.175\, \mathrm{km})$, and (c)
$(4.550\, \mathrm{km},\
1.738\, \mathrm{km})$.}\label{marginals}
\end{figure*}

To illustrate how the posterior regularized by the deep prior is
informed by the likelihood, we also calculated histograms at three
locations denoted by the white circles in Figure~\ref{std-colored}.
Histograms from the prior are calculated by randomly sampling network
weights from the prior distribution, i.e.,
$\mathrm{N}(\B{w} \mid \B{0}, \lambda^{-2}\B{I})$, followed by computing
the deep prior network's output for a random but fixed $\B{z}$. The
resulting histograms are plotted in light gray in
Figure~\ref{marginals}. Similarly, histograms for the posterior are
computed (equation~\ref{push-forward}) from samples of the posterior for
the weights. These are plotted in dark gray. As expected, the histograms
for the posterior are considerably narrower than those of the prior,
which means that the posterior is informed by the shot data. We also see
that the width of the histograms increases in areas with larger
variability. For comparison, we added the conditional mean estimates
with dashed vertical line. When compared with the ground truth values
denoted by the solid vertical lines, we observe that the ground truth
falls inside of the nonzero pointwise posterior interval, which confirms
the benefits of the prior.

\subsection{Accuracy and convergence
verification}\label{accuracy-and-convergence-verification}

Drawing samples from the posterior distribution via Markov chains can be
subject to errors when the chain is not long enough
\citep{gelman2013bayesian}. Unfortunately, the required length of the
chain is often infeasible in practice, certainly when the forward
operators is expensive to calculate as is the case with our imaging
examples. As explained earlier, we qualitative verify the accuracy of
the Bayesian inversion by comparing MAP estimates with confidence
intervals \citep{friedman2001elements} and by running different Markov
chains \citep{fang2018uqfip}.

\subsubsection{Consistency with empirical confidence
intervals}\label{consistency-with-empirical-confidence-intervals}

\begin{figure*}
\centering
\subfloat[\label{error-log-colored}]{\includegraphics[width=0.750\hsize]{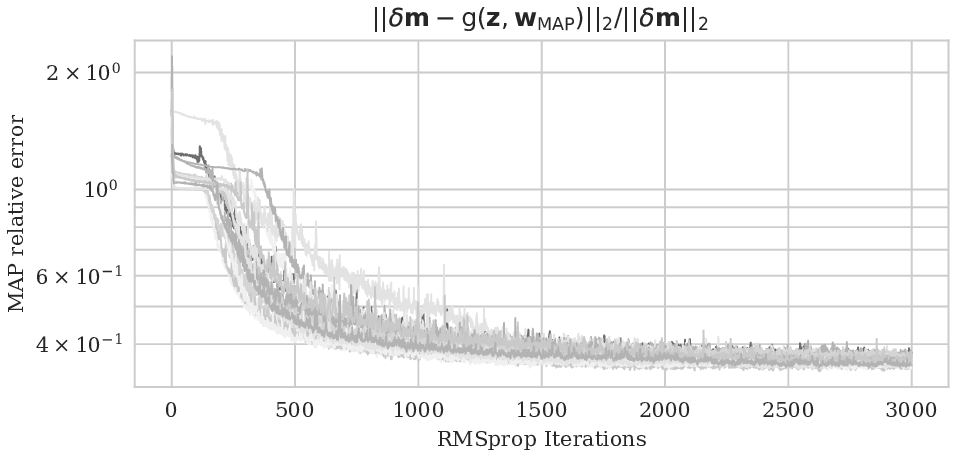}}
\\
\subfloat[\label{trace-80-colored}]{\includegraphics[width=0.350\hsize]{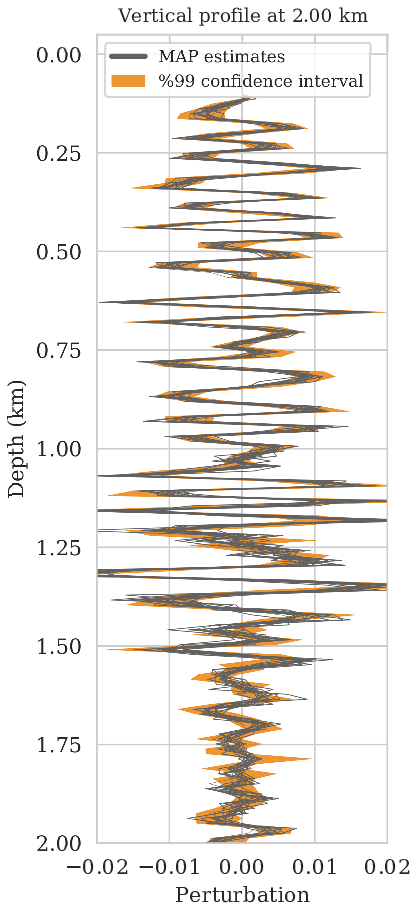}}
\subfloat[\label{trace-160-colored}]{\includegraphics[width=0.350\hsize]{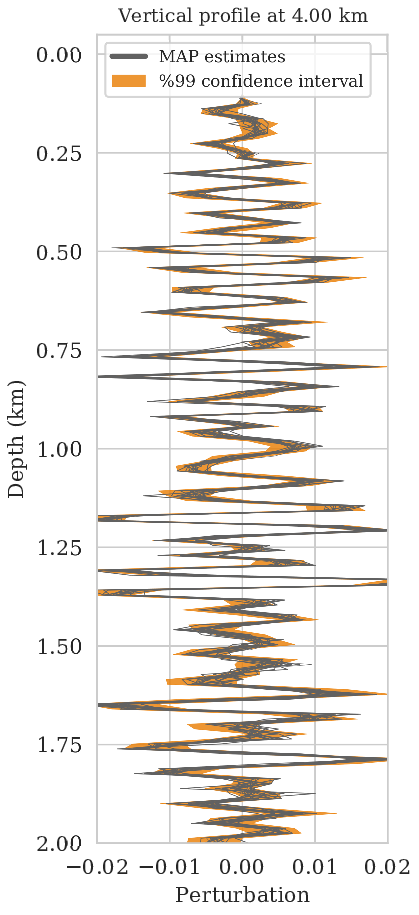}}
\caption{Confidence intervals empirical verification. (a) Relative error
in the estimated perturbation model for $15$ different initialization of
the deep prior, with respect to the ground truth image. Traces of
$99 \%$ confidence interval and $15$ realizations of the MAP estimate,
$\delta \B{m}_{\text{MAP}}$, at (b) $2.0\, \mathrm{km}$ and (c)
$4.0\, \mathrm{km}$ horizontal location.}\label{confidence-intervals}
\end{figure*}

As a first assessment of the accuracy of the MCMC sampling, we computed
the relative errors of $15$ MAP estimates with respect to the ground
truth image obtained for a single fixed $\B{z}$ but different
initializations of the network weights. The decay of the relative
$\ell_2$-norm error for each run over $3\mathrm{k}$ iterations are
plotted in Figure~\ref{error-log-colored} and show relatively small
variations from random realization to random realization. Vertical
profiles of the MAP estimates at two lateral positions confirm this
behavior. With few exceptions, these different MAP estimates fall well
within the shaded $99\%$ confidence intervals plotted in
Figures~\ref{trace-80-colored} and~\ref{trace-160-colored}. The
confidence intervals themselves were derived from samples of the
posterior. Except for perhaps the deeper part of the model, we can be
confident that the Bayesian inference is reasonable certainly in the
light of the nonlinearity of the deep prior itself.

\subsubsection{Chain to chain
variations}\label{chain-to-chain-variations}

\begin{figure*}
\centering
\subfloat[\label{mean-80-colored}]{\includegraphics[width=0.350\hsize]{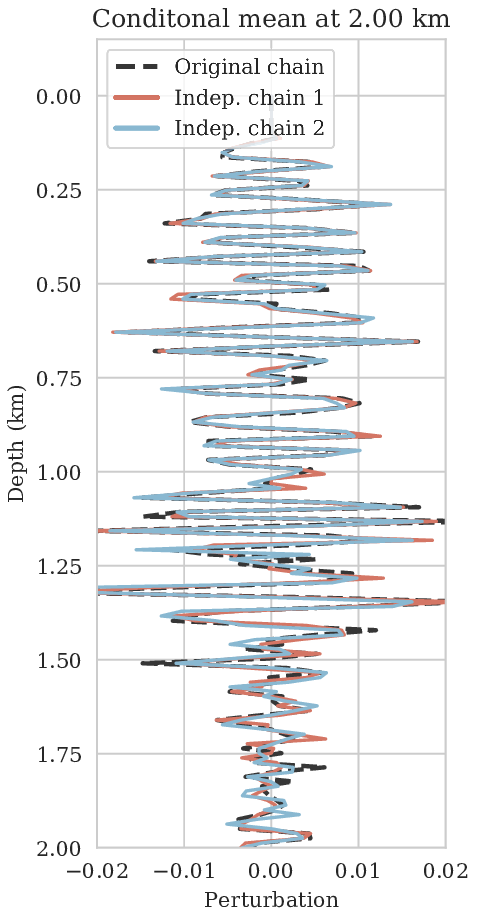}}
\subfloat[\label{mean-160-colored}]{\includegraphics[width=0.350\hsize]{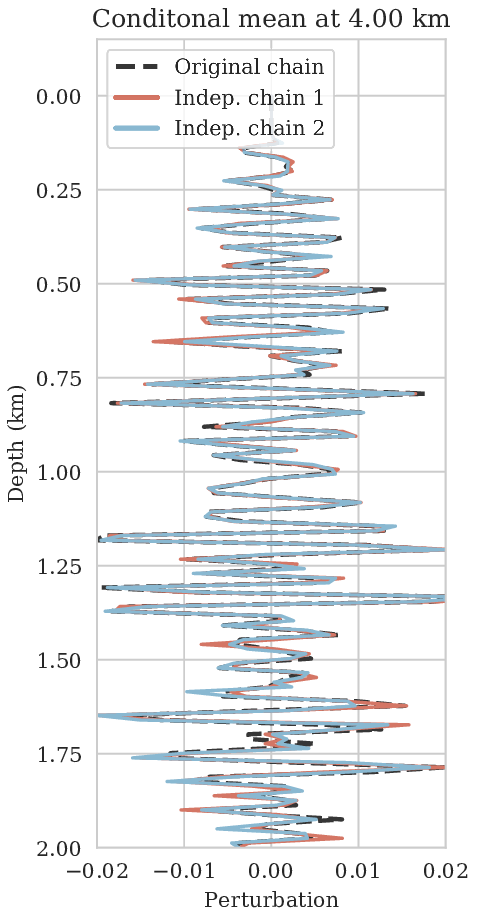}}
\\
\subfloat[\label{std-80-colored}]{\includegraphics[width=0.350\hsize]{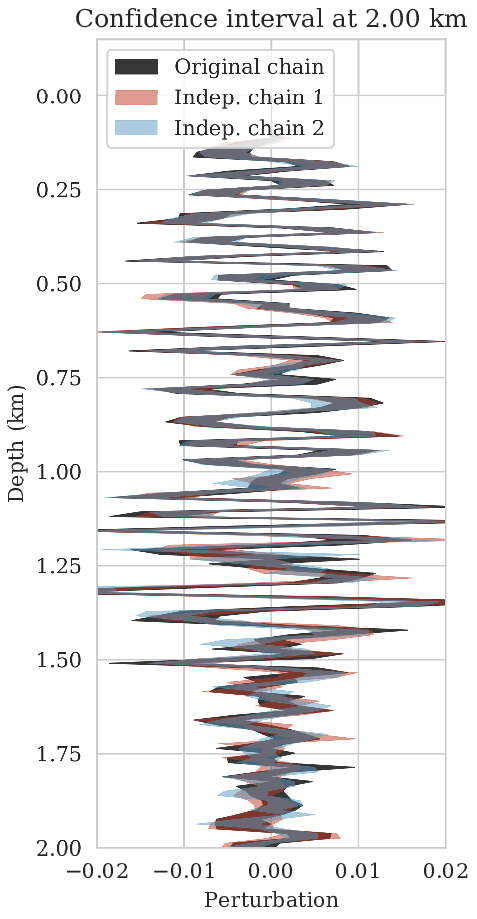}}
\subfloat[\label{std-160-colored}]{\includegraphics[width=0.350\hsize]{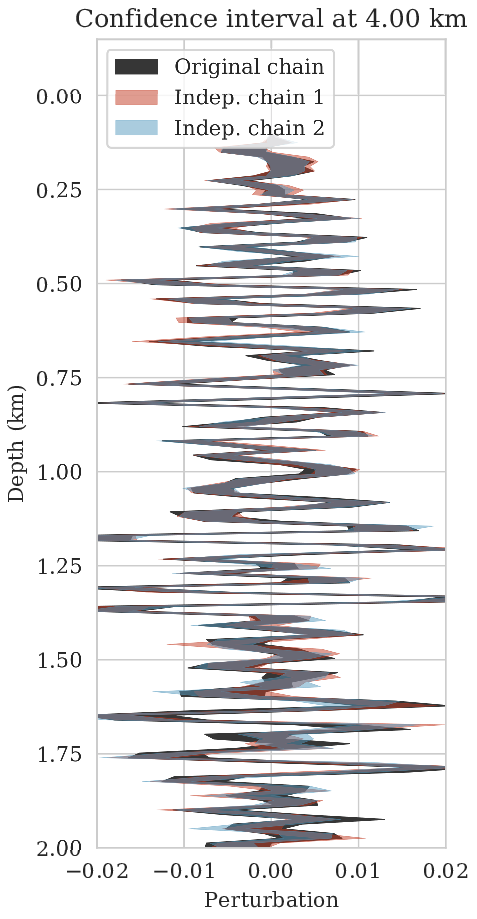}}
\caption{MCMC convergence diagnosis. Conditional posterior mean with
three independent Markov chains at (a) $2.0\, \mathrm{km}$ and (b)
$4.0\, \mathrm{km}$. Confidence intervals at (c) $2.0\, \mathrm{km}$ and
(d) $4.0\, \mathrm{km}$.}\label{mcmc-convergence}
\end{figure*}

To further assure our Bayesian inference is accurate, we conducted a
second experiment comparing estimates for the conditional mean and
confidence intervals for three different Markov chains computed from
independent random initialization for the weights and $\B{z}$ fixed.
Since we can not afford to run the Markov chains to convergence, we
expect slightly different results for the conditional mean and
confidence intervals. As observed from Figure~\ref{mcmc-convergence},
this is indeed the case but the variations are relatively minor and
confined to the deeper part of the image. This qualitative observation,
in conjunction with the behavior of the MAP estimates, suggests that we
can be confident that the presented Bayesian inference is reasonably
accurate certainly given the task of horizon tracking at hand.

\section{Probabilistic horizon
tracking}\label{probabilistic-horizon-tracking}

Typically, seismic images serve as input to a decision process involving
identification of certain attributes within the image preferably
including an assessment of their uncertainty. With very few exceptions
\citep{Ely2018}, these assessments of risk are not based on a systematic
approach where errors in shot data are propagated to uncertainty in the
image and subsequent tasks. To illustrate how the proposed Bayesian
inference can serve to assess uncertainty on downstream tasks, we
consider horizon tracking, where reflector horizons are extracted
automatically from seismic images given a limited number of user
specified control points. Typically, these control points are either
derived from well data available in the area or from human
interpretation. The horizon tracking can be deterministic, i.e.,
horizons are determined uniquely given a seismic image, or more
realistically, it may be nondeterministic, i.e., multiple possible
horizons explaining a single image. In either case, the task of
delineating the stratigraphy automatically with no to little
intervention by interpreters is challenging certainly in areas where the
geology is complex, e.g., near faults. To resolve these complex areas
high quality images including information on uncertainty are essential.

To set the stage to put tasks conducted on seismic images on a firm
statistical footing, we will make the assumption that these tasks are
only informed by the estimated image and not by the shot data
explicitly. This means when given an estimated image the task, e.g., of
horizon tracking, is assumed to be statistically independent of the shot
data. Formally, this conditional independence can be expressed as
\begin{equation}
( \B{h} \mathrel{\text{$\perp\mkern-10mu \perp$}}
    \B{d})  \mid \delta \B{m},
\label{conditional-independence}
\end{equation}
 where the random variable $\B{h}$ encodes tracked horizons. The symbol
$\mathrel{\text{$\perp\mkern-10mu \perp$}}$ represents conditional
independence \citep{Dawid1979} in this case between the estimated
horizons and shot records, given the seismic image, i.e., the seismic
images, $\delta \B{m}$, obtained from the shot records, $\B{d}$, contain
all the needed information to predict horizons, $\B{h}$. The assumed
statistical independence implies that the tracked horizons, $\B{h}$, can
be predicted unequivocally from estimated images. Because of the
independence, shot data does not bring forth additional information on
the horizons. For the remainder of this paper, we denote the task on the
image by $\mathcal{H}$, which for horizon tracking implies
$\B{h}=\mathcal{H}(\delta\B{m})$.

\subsection{Bayesian formulation}\label{bayesian-formulation}

Given the mapping from image to horizons, let
$p_{\mathcal{H}} \left (\B{h} \mid \delta \B{m} \right )$ represent the
conditional PDF of horizons given an estimate for the seismic image.
This distribution is characterized by the nondeterministic behavior of
$\mathcal{H}$ and assigns probabilities to horizons in the image,
$\delta \B{m}$. In the case where $\mathcal{H}$ represents automatic
horizon tracking this mapping requires control points as an additional
input. In this context, sampling from the conditional distribution is
equivalent to performing automatic horizon tracking with different
realizations of the control points. Alternatively, when $\mathcal{H}$
represents actions by human interpreters, samples from
$p_{\mathcal{H}} \left (\B{h} \mid \delta \B{m} \right )$ can be thought
of as horizons tracked by different individuals.

Provided samples from
$p_{\mathcal{H}} \left (\B{h} \mid \delta \B{m} \right )$ and assuming
conditional independence between $\B{h}$ and $\B{d}$ given the seismic
image described in equation~\ref{conditional-independence}, we can
perform Bayesian inference with the posterior distribution of horizons,
denoted by $p_{\text{post}} \left (\B{h} \mid \B{d} \right )$. Generally
speaking, for any arbitrary function of horizons, $f$, expectations with
respect to $p_{\text{post}} \left (\B{h} \mid \B{d} \right )$ can be
computed as follows:
\begin{equation}
\begin{aligned}
\mathbb{E}_{\B{h} \sim p_{\text{post}} \left (\B{h} \mid \B{d} \right )}
    \ \left [ f \left (\B{h} \right) \right ] & = \int f \left (
    \B{h} \right) p_{\text{post}} \left (\B{h} \mid \B{d} \right )
    \mathrm{d} \B{h}
= \iint f \left (\B{h} \right) p \left (\B{h}, \delta \B{m} \mid \B{d} \right )
    \mathrm{d} \B{h}\, \mathrm{d} \delta \B{m} \\
& = \iint f \left (\B{h} \right) p_{\mathcal{H}}
    \left (\B{h} \mid \delta \B{m}, \B{d} \right )
    p_{\text{post}} \left ( \delta
    \B{m} \mid \B{d} \right )
    \mathrm{d} \B{h}\, \mathrm{d} \delta \B{m} \\
& = \iint f \left (\B{h} \right) p_{\mathcal{H}}
    \left (\B{h} \mid \delta \B{m} \right )
    p_{\text{post}} \left ( \delta
    \B{m} \mid \B{d} \right )
    \mathrm{d} \B{h}\, \mathrm{d} \delta \B{m} \\
& = \mathbb{E}_{\delta \B{m} \sim p_{\text{post}} \left (\delta \B{m} \mid
    \B{d} \right ) } \left [ \int f \left (\B{h} \right)
    p_{\mathcal{H}} \left (\B{h} \mid \delta \B{m} \right )
     \mathrm{d} \B{h} \right ] \\
& = \underbrace {\mathbb{E}_{\delta \B{m} \sim p_{\text{post}} \left (\delta \B{m} \mid
    \B{d} \right ) } \underbrace {\mathbb{E}_{
    \B{h} \sim p_{\mathcal{H}} \left (\B{h} \mid \delta \B{m} \right ) } \left [
    f \left (\B{h} \right) \right ]. }_{\substack{
    \text{ uncertainty in} \\ \text{ horizon tracking}}}}_{
    \substack{\text{ uncertainty in} \text{ seismic imaging}}} \\
\end{aligned}
\label{horizon-inference}
\end{equation}
 The second equality in the first line of
equation~\ref{horizon-inference} follows from the law of total
probability\footnote{$p(x) = \int_{\mathcal{Y}} p(x, y)\,\mathrm{d}y$,
  where $x \in \mathcal{X}$ and $y \in \mathcal{Y}$ are two arbitrary
  random variables.}, the second line is obtained by applying the chain
rule of PDFs\footnote{$p(x, y) = p(x \mid y)\, p(y), \ \forall\, x \in \mathcal{X},\, y \in \mathcal{Y}$.}
to the joint density $p \left (\B{h}, \delta \B{m} \mid \B{d} \right )$,
and the third line exploits the conditional independence assumption in
equation~\ref{conditional-independence}. Conceptually,
equation~\ref{horizon-inference} states that we can decompose the
uncertainty in horizon tracking into two parts, namely uncertainty in
imaging and uncertainty in the horizon tracking task itself. Based on
equation~\ref{horizon-inference}, expectations over
$p_{\text{post}} \left (\B{h} \mid \B{d} \right )$ can be calculated via
Monte Carlo integration using samples from
$p_{\text{post}} \left (\B{h} \mid \B{d} \right )$. Thanks to the
conditional independence assumption in
equation~\ref{conditional-independence}, we can sample from
$p_{\text{post}} \left (\B{h} \mid \B{d} \right )$ by sampling the
imaging posterior,
$p_{\text{post}} \left (\delta \B{m} \mid \B{d} \right )$, followed by
tracking the horizons in each seismic image. This step yields an
ensemble of possible horizons for each sampled image. Using samples
drawn from $p_{\text{post}} \left (\B{h} \mid \B{d} \right )$, we
approximate the expectation in equation~\ref{horizon-inference} by the
sample mean. In the following sections, we break
equation~\ref{horizon-inference} down into two cases where the horizon
tracker yields an unique set of horizons or multiple sets of likely
horizons, given one seismic image.

\subsubsection{\texorpdfstring{Case $1$: horizons are unique given an
image}{Case 1: horizons are unique given an image}}\label{case-1-horizons-are-unique-given-an-image}

In the simplest case, where horizon tracking uniquely determines the
horizons given a seismic image, the conditional PDF
$p_{\mathcal{H}} \left (\B{h} \mid \delta \B{m} \right )$ corresponds to
a delta function, i.e., we have
\begin{equation}
\begin{split}
p_{\mathcal{H}}(\B{h} \mid \delta \B{m}) = \delta_{\left [
    \B{h} = \mathcal{H} \left ( \delta \B{m} \right )
    \right ]} \left (\B{h} \right ), \\
\end{split}
\label{probH}
\end{equation}
 where $\mathcal{H}$ represent the deterministic horizon tracking map
and $\delta(\cdot)$ stands for the delta Dirac distribution.
Substituting equation~\ref{probH} into equation~\ref{horizon-inference}
yields
\begin{equation}
\begin{aligned}
\mathbb{E}_{\B{h} \sim p_{\text{post}} \left (\B{h} \mid \B{d} \right )}
    \ \left [ f \left (\B{h} \right) \right ]
& = \mathbb{E}_{\delta \B{m} \sim p_{\text{post}} \left (\delta \B{m} \mid
    \B{d} \right ) } \left [ \int f \left (\B{h} \right)
    \delta_{\left [\B{h} = \mathcal{H} \left ( \delta \B{m}
    \right ) \right ]} \left (\B{h} \right ) \mathrm{d}
    \B{h} \right ], \\
& = \mathbb{E}_{\delta \B{m} \sim p_{\text{post}} \left (\delta \B{m} \mid
    \B{d} \right ) } \left [ f \left ( \mathcal{H} \left ( \delta
    \B{m} \right ) \right ) \right ], \\
& \approx \frac{1}{n_{\mathrm{w}}}\sum_{j=1}^{n_{\mathrm{w}}} f
    \left ( \mathcal{H} ( {\delta \B{m}}_j ) \right ), \\
\end{aligned}
\label{independence-det}
\end{equation}
 where
$\left \{ {\delta \B{m}}_j \right \}_{j=1}^{n_{\mathrm{w}}} \sim p_{\text{post}} ( \delta \B{m} \mid \B{d} )$
are $n_{\mathrm{w}}$ samples from the posterior distribution.
Equation~\ref{independence-det} essentially means that in case of a
deterministic horizon tracker uncertainty in imaging can be translated
to uncertainty in horizon tracking by simply drawing samples from the
seismic imaging posterior and tracking horizons in each image. This
procedure results in samples from the posterior distribution of horizons
and inference of this posterior distribution is done via the equation
above.

\subsubsection{\texorpdfstring{Case $2$: multiple likely horizons given
an
image}{Case 2: multiple likely horizons given an image}}\label{case-2-multiple-likely-horizons-given-an-image}

The probabilistic horizon tracking approach, described in
equation~\ref{horizon-inference}, also admits nondeterministic horizon
trackers, e.g., automatic horizon tracking with uncertain control
points, e.g., points provided by multiple human interpreters. In this
case, instead of having one set of horizons for each seismic image, we
have multiple realizations of horizons that each agree with a seismic
image, i.e., they are samples from
$p_{\mathcal{H}}(\B{h} \mid \delta \B{m})$. With these samples, the
inner expectation in equation~\ref{horizon-inference} can be estimated.
Assuming that for each image we have $n_h$ different realizations of
tracked horizons, namely,
$h_k^{(\delta \B{m})} \sim p_{\mathcal{H}}(\B{h} \mid \delta \B{m}), \
k=1,\cdots, n_h$, equation~\ref{horizon-inference} becomes
\begin{equation}
\begin{aligned}
\mathbb{E}_{\B{h} \sim p_{\text{post}} \left (\B{h} \mid \B{d} \right )}
    \ \left [ f \left (\B{h} \right) \right ]
& = \mathbb{E}_{\delta \B{m} \sim p_{\text{post}} \left (\delta \B{m} \mid
    \B{d} \right ) } \mathbb{E}_{
    \B{h} \sim p_{\mathcal{H}} \left (\B{h} \mid \delta \B{m} \right ) } \left [
    f \left (\B{h} \right) \right ] \\
& \approx \mathbb{E}_{\delta \B{m} \sim p_{\text{post}} \left (\delta \B{m} \mid
    \B{d} \right ) } \left [ \frac{1}{n_h} \sum_{k=1}^{n_h}
    f \left ( h_k^{(\delta \B{m})}\right )
    \right ], \\
& \approx \frac{1}{n_h n_{\mathrm{w}}} \sum_{j=1}^{n_{\mathrm{w}}}
    \sum_{k=1}^{n_h} f \left ( h_k^{({\delta \B{m}}_j)}\right )
\end{aligned}
\label{sum-horizons-points}
\end{equation}
 where $h_k^{({\delta \B{m}}_j)}$ is the $k$th sample from
$p_{\mathcal{H}}(\B{h} \mid {\delta \B{m}}_j)$ and ${\delta \B{m}}_j$ is
the $j$th sample from
$p_{\text{post}} \left (\delta \B{m} \mid \B{d} \right)$. Because it has
an extra sum over the different realizations of tracked horizons for a
fixed image, equation~\ref{sum-horizons-points} differs from
equation~\ref{independence-det}. In the ensuing sections, we show how
equations~\ref{independence-det} and~\ref{sum-horizons-points} can be
used to calculate pointwise estimates for the first two moments of the
posterior distribution over horizons.

\subsubsection{Uncertainty quantification in horizon
tracking}\label{uncertainty-quantification-in-horizon-tracking}

It is often beneficial to express uncertainty in the form of confidence
intervals. For this purpose, equation~\ref{independence-det}
or~\ref{sum-horizons-points} is evaluated first for $f(\B{h}) =\B{h}$.
This yields the conditional mean estimate for the horizons denoted by
\begin{equation}
\boldsymbol{\mu}_{\B{h}} = \mathbb{E}_{\B{h} \sim p_{\text{post}} \left (\B{h} \mid \B{d} \right )} [ \B{h} ].
\label{cm-horizons}
\end{equation}
 Similarly, the pointwise standard deviation of the horizons can be
computed by choosing
$f(\B{h}) = (\B{h} - \boldsymbol{\mu}_{\B{h}}) \odot (\B{h} - \boldsymbol{\mu}_{\B{h}})$.
This latter point estimate can be calculated via
\begin{equation}
\boldsymbol{\sigma}^2_{\B{h}} = \mathbb{E}_{\B{h} \sim p_{\text{post}} \left (\B{h} \mid \B{d} \right )} [ (\B{h} - \boldsymbol{\mu}_{\B{h}}) \odot (\B{h} - \boldsymbol{\mu}_{\B{h}}) ],
\label{std-horizons}
\end{equation}
 where $\boldsymbol{\sigma}_{\B{h}}$ denotes the pointwise standard
deviation. The $99\%$ confidence interval for horizons is the interval
$\boldsymbol{\mu}_{\B{h}} \pm 2.576\, \boldsymbol{\sigma}_{\B{h}}$.
Contrary to most existing automatic horizon trackers, the uncertainty
estimates we provide here are determined by uncertainties in the image
due to noise, and possibly linearization errors, in the shot records.

\section{\texorpdfstring{Probabilistic horizon tracking---``ideal
low-noise'' Parihaka
example}{Probabilistic horizon tracking---ideal low-noise Parihaka example}}\label{probabilistic-horizon-trackingideal-low-noise-parihaka-example}

The main goal of this paper is to derive a systematic approach to
propagate uncertainty in imaging to the task at hand. For this purpose,
we first consider the relatively ideal case of uncertainty due to
additive random noise in the shot data. To illustrate how the presence
of this noise affects the task of horizon tracking, we apply the
proposed probabilistic framework to the Parihaka imaging example
discussed earlier. Because the seismic shot data for this example is
relatively low-frequency ($30\,$Hz source peak frequency) and the
geology relatively simple, horizons are not that challenging to track.
However, there is a substantial amount of noise in the shot data that we
need to contend with when tracking the imaged horizons. For the latter
task, we deploy the tracking approach introduced by \citet{wu2018least},
which requires the user to provide control points on the seismic
horizons of interest.

\begin{figure*}
\centering
\includegraphics[width=0.700\hsize]{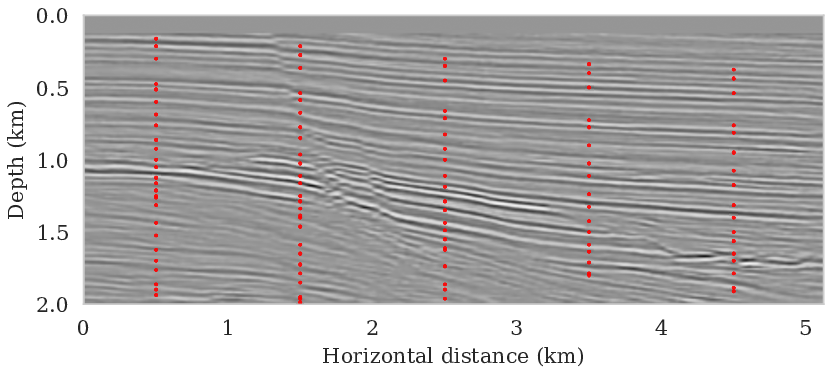}
\caption{Five sets of control points identifying $25$ horizons of
interest.}\label{control-points}
\end{figure*}

To setup this tasked imaging experiment, we select $25$ horizons from
the conditional mean estimate (Figure~\ref{cm-colored}) calculated for
the Parihaka seismic imaging example. Next, control points are picked
for the selected horizons at various horizontal positions, separated by
$1\, \mathrm{km}$. We group the control points with the same horizontal
location, yielding five sets of control points.
Figure~\ref{control-points} shows these five sets located at lateral
positions $0.5$, $1.5$, $2.5$, $3.5$, and $4.5\, \mathrm{km}$.

To separate the effect of errors in the shot data and variations amongst
provided control points, we consider noisy shot data first, followed by
the situation where there is uncertainty due to noise in the shot data
and due to variations in the control points.

\subsection{Uncertainty due to noise in shot data (case
1)}\label{uncertainty-due-to-noise-in-shot-data-case-1}

\begin{figure*}
\centering
\subfloat[\label{cp-40-colored}]{\includegraphics[width=0.500\hsize]{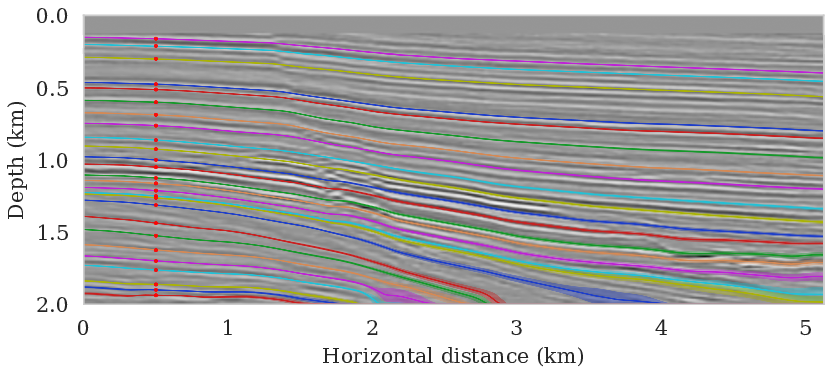}}
\subfloat[\label{cp-80-colored}]{\includegraphics[width=0.500\hsize]{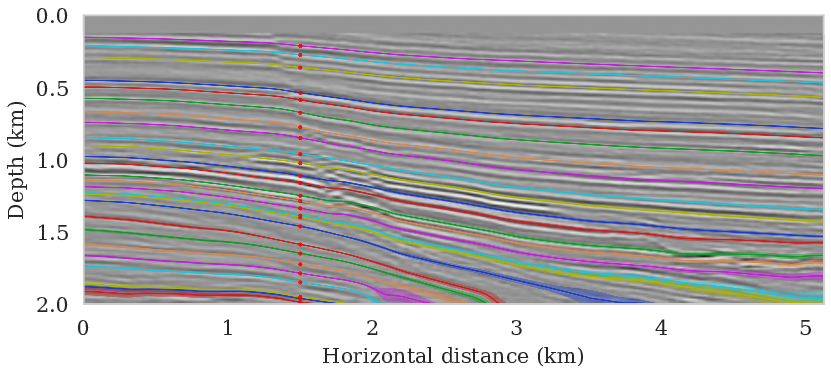}}
\\
\subfloat[\label{cp-120-colored}]{\includegraphics[width=0.500\hsize]{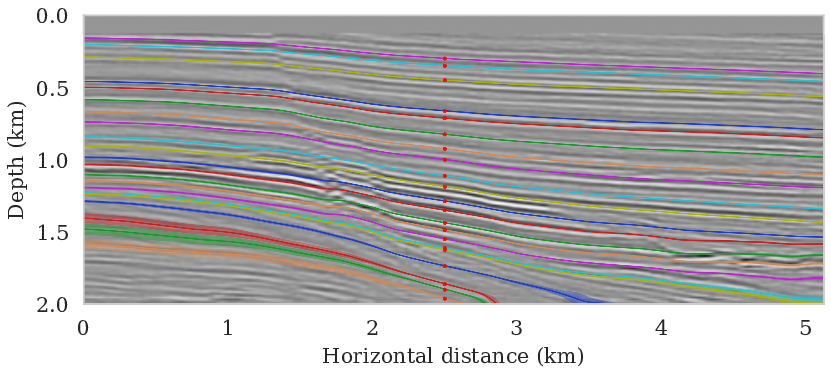}}
\subfloat[\label{cp-160-colored}]{\includegraphics[width=0.500\hsize]{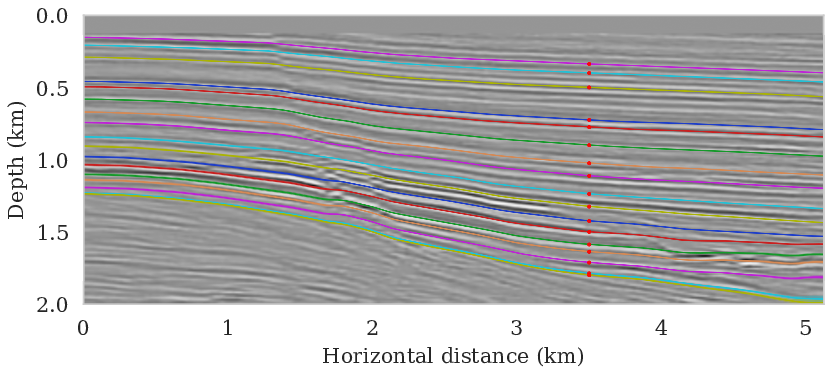}}
\\
\subfloat[\label{cp-200-colored}]{\includegraphics[width=0.500\hsize]{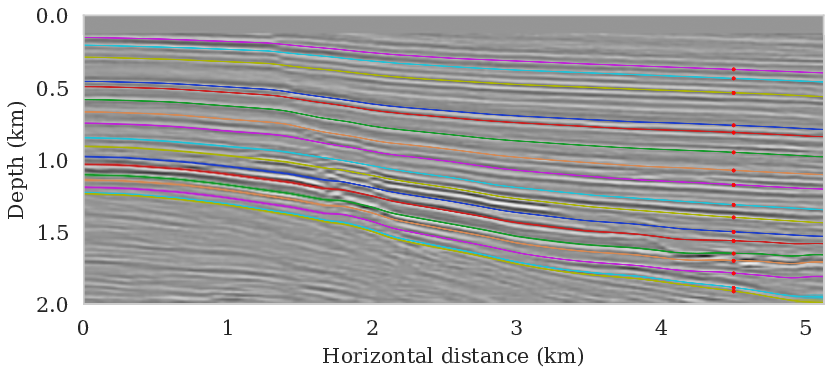}}
\caption{Uncertainty in horizon tracking only due noise in the shot data
(equation~\ref{independence-det}). Control points are located at (a)
$500\, \mathrm{m}$, (b) $1500\, \mathrm{m}$, (c) $2500\, \mathrm{m}$,
(d) $3500\, \mathrm{m}$, and (e) $4500\, \mathrm{m}$ horizontal
location. Conditional mean estimates and $99\%$ confidence intervals are
shown in solid and shaded colors, respectively.}\label{horizon-uq-det}
\end{figure*}

To calculate noise-induced uncertainties in horizon tracking (case 1),
we pass samples from the imaging posterior distribution,
$p_{\text{post}} \left (\delta \B{m} \mid \B{d} \right )$, to the
automatic horizon tracking software \citep{wu2018least}. Given the five
sets of selected control points (Figure~\ref{control-points}), the
tracker generates for each sample of the imaging posterior $25$ horizons
according to equation~\ref{independence-det}. For each set of control
points, the conditional mean and $99\%$ confidence intervals are
calculated included in Figure~\ref{horizon-uq-det}. Each plot
(Figures~\ref{cp-40-colored} --~\ref{cp-200-colored}) corresponds to
tracked horizons with confidence intervals derived from different sets
of control points. As expected, the results exhibit more uncertainty for
horizons tracked in the deeper parts of the image and close to
boundaries, which is consistent with the relative poor illumination in
these areas. Moreover, uncertainty in the tracked locations increases
away from the control points. This increase in uncertainty agrees with
the inherent challenge of automatic horizon tracking across areas of
poor illumination, faults and tortuous reflectors. This behavior is
observed for each set of control points.

Aside from shot noise induced uncertainty, variations in the control
point may also contribute to uncertainty in the horizon tracking. This
corresponds to case 2, which we consider in the next section.

\subsection{Uncertainty due to noise and uncertain control points (case
2)}\label{uncertainty-due-to-noise-and-uncertain-control-points-case-2}

The presence of noise in shot data is often not the only cause of
uncertainty within the task of (automatic) horizon tracking. Human
errors, or better variations in the selection of control points by
interpreters, may also contribute to uncertainty. To mimic differences
in selected control points, we impose a distribution over the control
points. For simplicity, in this example we assume that the five sets of
control points are equally likely to be accurate. This is to say, we are
equally certain of the accuracy of the picked control points. This can
be related to the case where we have access to wells in the seismic
survey area and we are certain of the well tying procedure. Other
sources of error such as uncertainty in location of the control points,
multiple human interpreters, etc, can also be incorporated in the
equation~\ref{sum-horizons-points}, but will not be considered here.
Given the above assumption on the probability distribution, each
realization of the seismic image gives rise to multiple equally likely
tracked horizons for each of the five sets of control points.

\begin{figure*}
\centering
\includegraphics[width=0.750\hsize]{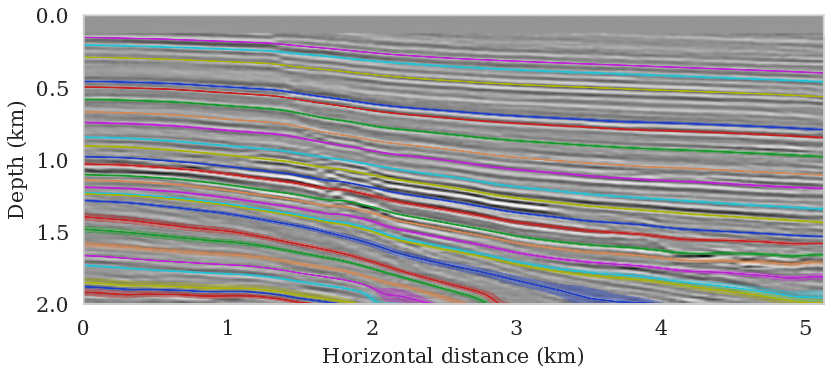}
\caption{Uncertainty in horizon tracking due to a combination of
uncertainties in imaging and control points
(equation~\ref{sum-horizons-points}).}\label{horizon-uq-nondet}
\end{figure*}

Given these multiple tracked horizons for each image, pointwise first
and second moments of
$p_{\mathcal{H}} \left (\B{h} \mid \delta \B{m} \right )$ are calculated
via equations~\ref{cm-horizons} and~\ref{std-horizons} by tracking
horizons in each sample from the imaging posterior. The horizon tracker
yields five sets of horizons for each seismic image, each obtained using
one of the five sets of control points. Results of this procedure are
summarized in Figure~\ref{horizon-uq-nondet}. As in the earlier
examples, we observe an increase in uncertainty with depth and at the
boundaries. Contrary to small uncertainties near the control points, we
now observe uncertainty everywhere along the tracked horizons, which
suggests increased variability amongst horizons. The variability comes
from not trusting only one set of control points, but incorporating
information from all the fives sets of control points.

\section{\texorpdfstring{Probabilistic horizon tracking---``noisy''
high-resolution Compass model
example}{Probabilistic horizon tracking---noisy high-resolution Compass model example}}\label{probabilistic-horizon-trackingnoisy-high-resolution-compass-model-example}

While the seismic imaging examples considered so far were directly
obtained from migrated shot data of the Parihaka dataset, the
linearization errors, i.e., errors due to linearizing the wave equation
with respect to the background squared-slowness model, were ignored
because the quasi-real data was generated through the demigration
process for a made-up background model. Aside from this simplification,
the geology of the examples discussed so far was relatively simple and
imaged at low resolution. To account for a more realistic setting,
involving complex geology and high-resolution imaging, we will quantify
imaging and horizon tracking uncertainty given high-frequency but noisy
synthetic shot data. To mimic the complexities of field data, noisy shot
data---generated with nonlinear forward modeling on a 2D subset of the
Compass model \citep{Jones2012}---is used as input to the proposed
imaging scheme. We select the synthetic Compass model because it
contains realistic heterogeneity derived from both seismic and well data
collected in the North Sea \citep{Jones2012}. Aside from a low
signal-to-noise ratio of $-9.17\, \mathrm{dB}$, this example is affected
by linearization errors.

As before, we first describe the problem setup, followed by a comparison
between the unregularized MLE and deep-prior regularized MAP and
conditional mean estimates. After presenting results on uncertainty
quantification on the image, results of our probabilistic horizon
tracking framework will be discussed below.

\subsection{Problem setup}\label{problem-setup-1}

Split spread raw data, consisting of $101$ shot records sampled with a
source spacing of $25\, \mathrm{m}$~and a receiver sampling
$12.5\, \mathrm{m}$, is generated by solving the acoustic wave equation
for the 2D subset of the Compass model. To mimic broadband data, the
source is a Ricker wavelet with a central frequency of
$40\, \mathrm{Hz}$. Each shot is $1.6$ seconds long.

To image the shot data, we first derive a kinematically correct
background model via smoothing. Next, linearized data is created by
subtracting shot data simulated in this smooth background model from the
shot data simulated in the actual model. This data serves as input to
our imaging scheme.

\subsection{Imaging with uncertainty
quantification}\label{imaging-with-uncertainty-quantification}

To arrive at the MLE and MAP estimates for the image, we follow the
procedure outlined before. The MLE estimate is obtained by minimizing
equation~\ref{deep-prior-likelihood} with stochastic optimization
limiting the number of iterations to six passes over the $101$ shots.
Equation~\ref{deep-prior} is minimized with respect to $\B{w}$ with the
RMSprop optimization algorithm \citep{rmsprop} for stepsize of $10^{-3}$
and five thousand iterations (about $50$ passes over all shots). We
stopped the iterations after no further visual improvement to the image
was observed. The value for the tradeoff off parameter,
$\lambda^{-2}=3 \times 10^{-5}$, was set after extensive parameter
testing guided by a value that leads to the least amount of visual
imaging artifacts.

\begin{figure*}
\centering
\subfloat[\label{BG_mle-colored}]{\includegraphics[width=0.700\hsize]{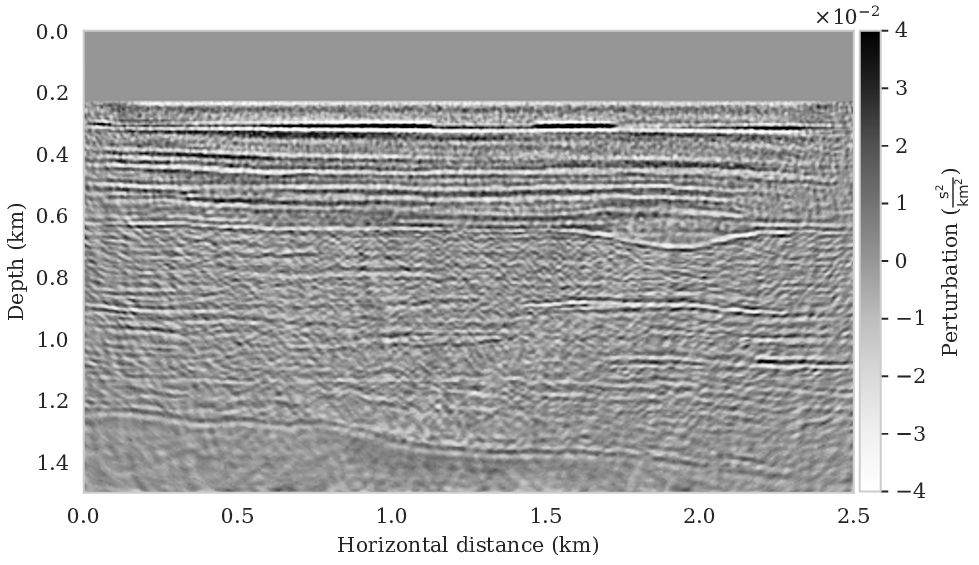}}
\\
\subfloat[\label{BG_map-colored}]{\includegraphics[width=0.700\hsize]{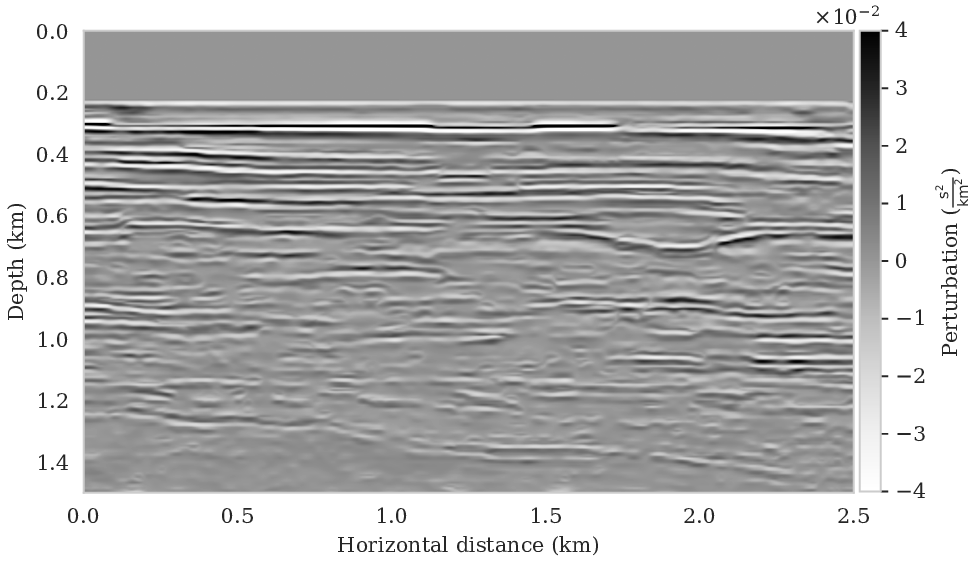}}
\\
\subfloat[\label{BG_cm-colored}]{\includegraphics[width=0.700\hsize]{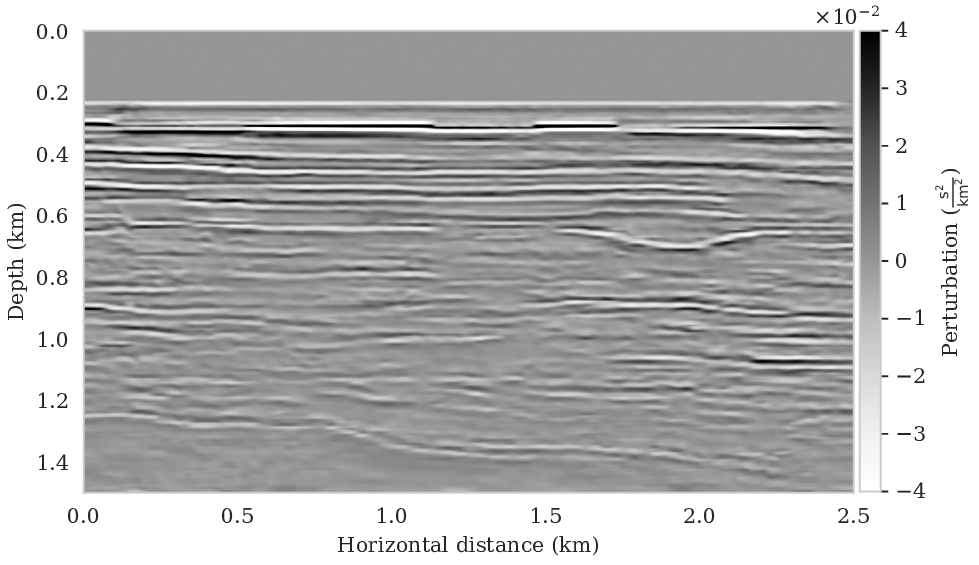}}
\caption{Imaging a 2D subset of the Compass model with deep priors. (a)
MLE, i.e., minimizer of equation~\ref{imaging-likelihood} with respect
to $\delta \B{m}$, with SNR $2.80\,$dB. (b) The MAP estimate, i.e.,
minimizer of equation~\ref{imaging-obj}, following by mapping onto the
image space via $g$ (equation~\ref{MAP}), with SNR $3.91\,$dB. (c) The
conditional (posterior) mean estimate, $\delta \B{m}_{\text{CM}}$, with
SNR $4.12\,$dB.}\label{BG_imaging-results}
\end{figure*}

Compared to the previous example, the MLE image estimate is of poor
quality (SNR $2.80\,$dB) and contains many imaging artifacts stemming
from the noise and linearization errors. The MAP estimate, on the other
hand, is improved (SNR $3.91\,\mathrm{dB}$) thanks to regularization by
the deep prior but it does contain unrealistic artifacts and misses
details especially in the deeper parts of the image (juxtapose
Figures~\ref{BG_mle-colored} and~\ref{BG_map-colored}). By running
Algorithm~\ref{alg} for ten thousand iterations, we compute five
thousand samples from the posterior distribution on the image following
the stepsize strategy of equation~\ref{stepsize} with
$\gamma = \frac{1}{3}$ and $a,\ b$ chosen such the stepsize decreases
from $5 \times 10^{-3}$ to $10^{-3}$. This took approximately $18$ hours
on a quad-core machine. The resulting estimate for the conditional mean,
included in Figure~\ref{BG_cm-colored}, represents a considerable visual
improvement with a SNR of $4.12\,$dB. Compared to the MLE and MAP
estimates, the conditional mean estimate exhibits more continuous
reflectors and significantly fewer artifacts. This example confirms that
images yielded by the conditional mean of inverse problems regularized
by the deep prior are relatively robust to noise.

\begin{figure*}
\centering
\subfloat[\label{BG_std-colored}]{\includegraphics[width=0.700\hsize]{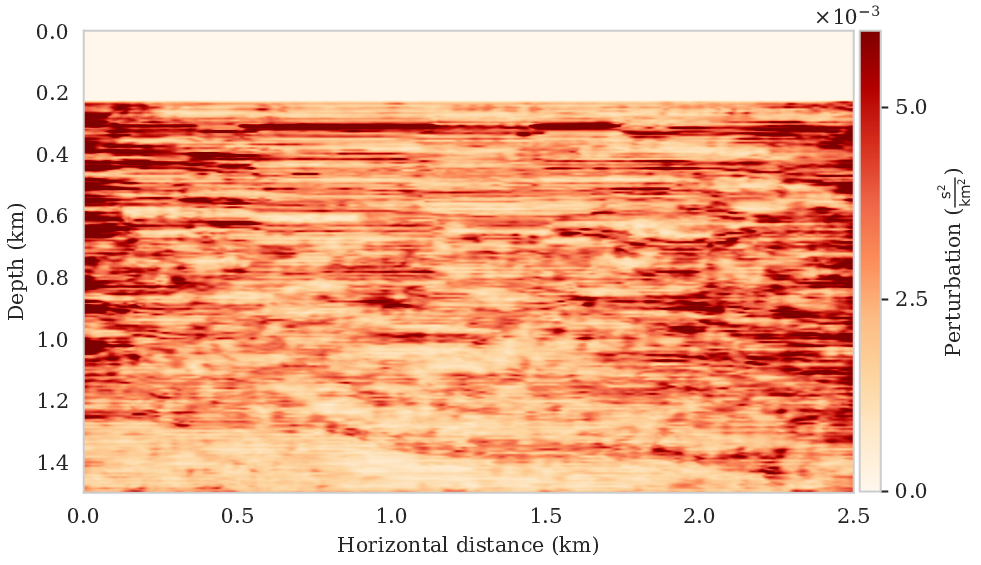}}
\\
\subfloat[\label{BG_std_normalized}]{\includegraphics[width=0.700\hsize]{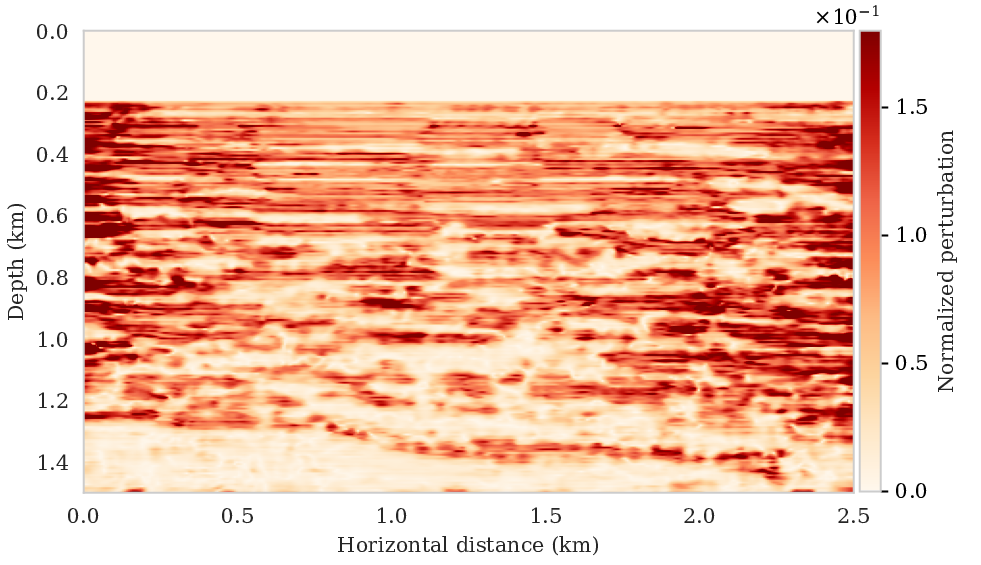}}
\caption{Imaging uncertainty quantification corresponding to the Compass
model. (a) The pointwise standard deviation among samples drawn from the
posterior, $\boldsymbol{\sigma}_{\text{post}}$. (b) Normalized pointwise
standard deviation by the conditional mean estimate
(Figure~\ref{BG_cm-colored}).}\label{BG_imaging-std}
\end{figure*}

The available samples from the posterior also allows us to calculate an
estimate for the pointwise standard deviation of the image. This
quantity is plotted in Figure~\ref{BG_std-colored}. To avoid overprint
by the strong reflectors, we also included a plot of the normalized
standard deviation obtained by stabilized division by the conditional
mean. Both plots for the pointwise standard deviations show a distinct
correlation between difficult to image areas of complex geology, such as
channels, and areas affected by relatively poor illumination near the
edges and for the deep parts of the image. In the next section, we will
show how the samples from the posterior inform the task of horizon
tracking.

\subsection{Horizon tracking with uncertainty
quantification}\label{horizon-tracking-with-uncertainty-quantification}

Similar to the previous horizon tracking example $14$ horizons are
selected from the conditional mean estimate for the image
(Figure~\ref{BG_cm-colored}). As before, control points are picked for
each horizon at the horizontal locations $62.5$, $562.5$, $1062.5$,
$1562.5$, and $2062.5\, \mathrm{m}$. Control points with the same
horizontal position are grouped together, yielding five sets of control
points. As before, we distinguish between uncertainties related to noise
and now also linearization errors in the data and uncertainty related to
errors in the control parameters and in the shot data due to noise and
linearization errors.

\subsubsection{Uncertainty due to noise and linearization errors in the
shot data (case
1)}\label{uncertainty-due-to-noise-and-linearization-errors-in-the-shot-data-case-1}

\begin{figure*}
\centering
\subfloat[\label{BG_cp-10-colored}]{\includegraphics[width=0.500\hsize]{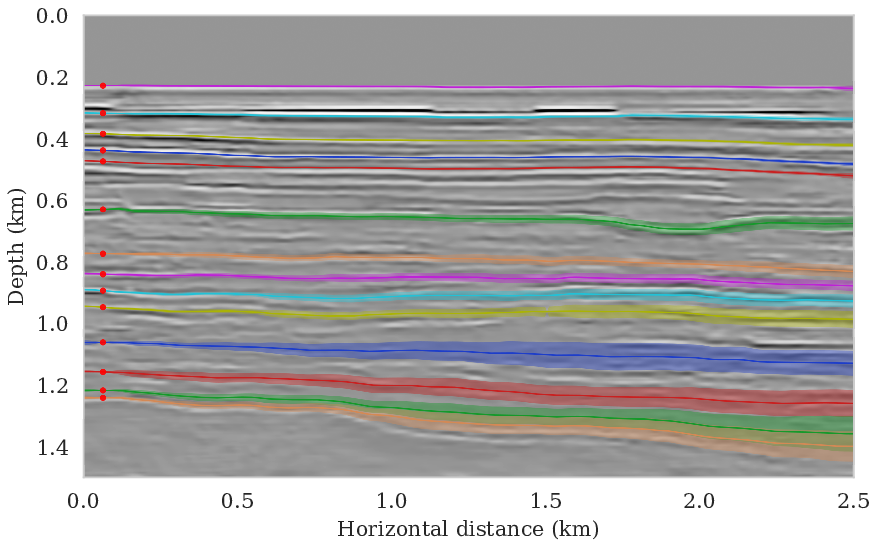}}
\subfloat[\label{BG_cp-90-colored}]{\includegraphics[width=0.500\hsize]{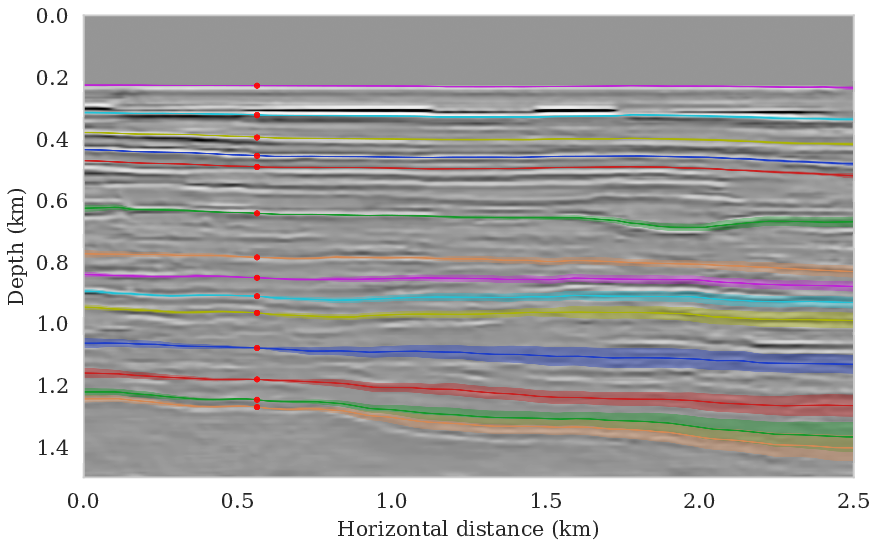}}
\\
\subfloat[\label{BG_cp-170-colored}]{\includegraphics[width=0.500\hsize]{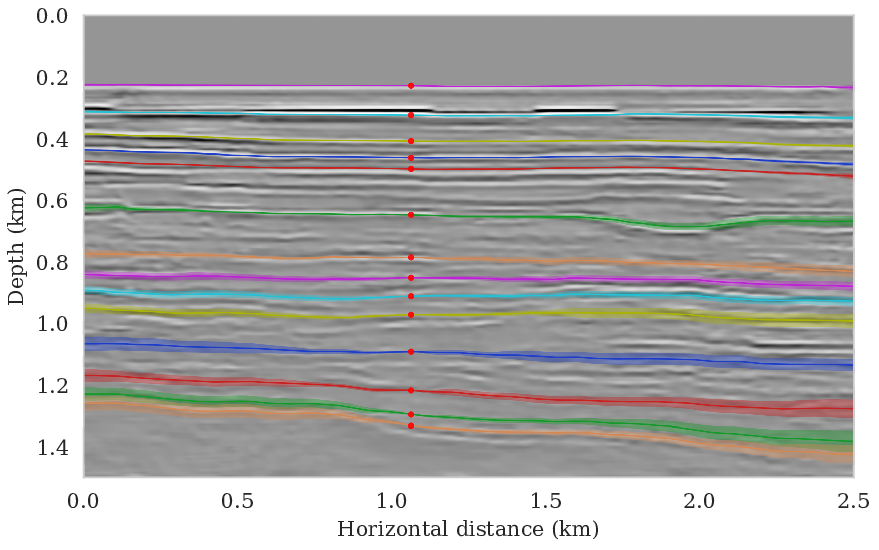}}
\subfloat[\label{BG_cp-250-colored}]{\includegraphics[width=0.500\hsize]{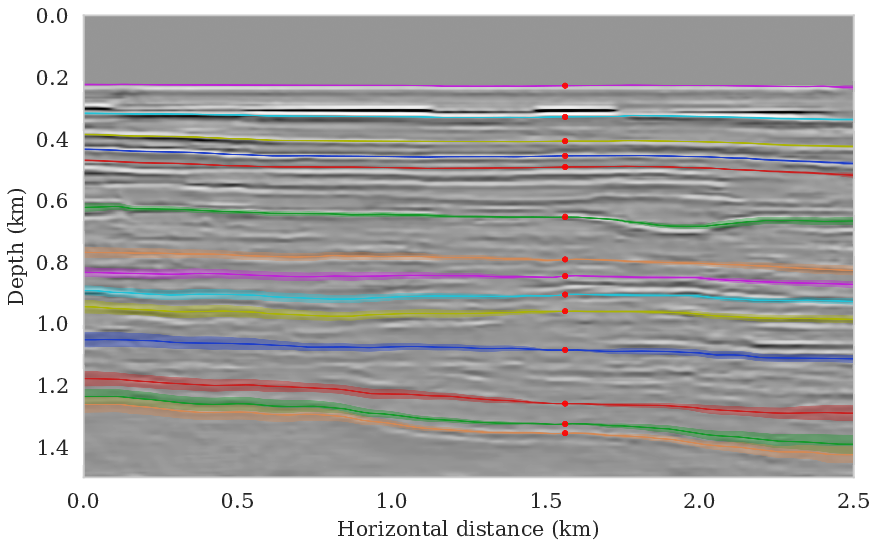}}
\\
\subfloat[\label{BG_cp-350-colored}]{\includegraphics[width=0.500\hsize]{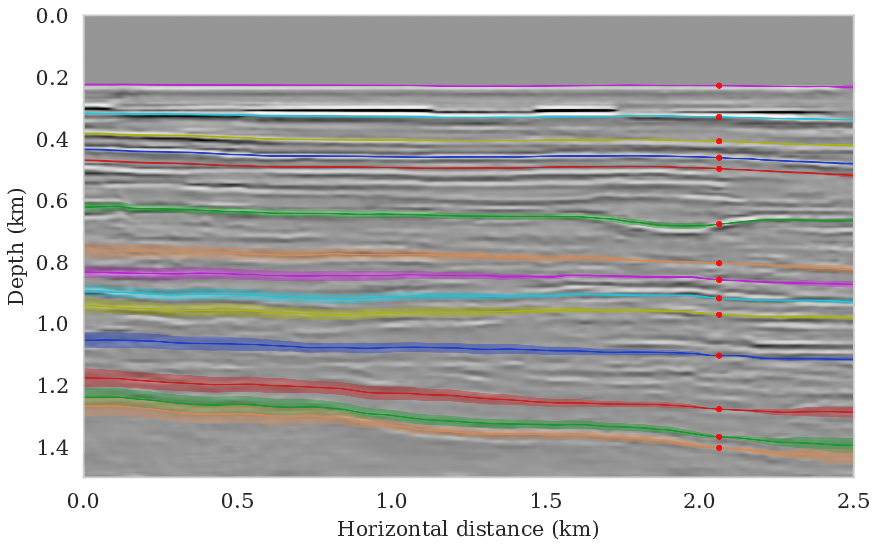}}
\caption{Uncertainty in horizon tracking due uncertainties in imaging
(equation~\ref{independence-det}). Control points are located at (a)
$62.5\, \mathrm{m}$, (b) $562.5\, \mathrm{m}$, (c)
$1062.5\, \mathrm{m}$, (d) $1562.5\, \mathrm{m}$, and (e)
$2062.5\, \mathrm{m}$ horizontal location. Conditional mean estimates
and $99\%$ confidence intervals are shown in solid and shaded colors,
respectively.}\label{BG_horizon-uq-det}
\end{figure*}

Tracked horizons plus $99\%$ confidence intervals for the five sets of
control points are included in Figures~\ref{BG_cp-10-colored}
--~\ref{BG_cp-350-colored}. These results were computed by sampling the
posterior distribution for horizon tracking,
$p_{\text{post}} \left (\B{h} \mid \B{d} \right)$, following the
procedure described above. Not unexpected, we consistently observe
increases in uncertainty as we move further away from the control points
and deeper into the image. This increase in the size of the confidence
interval is due to the increased variability amongst the samples of the
imaging posterior especially in regions that are more difficult to
image, e.g., near the boundaries of the image, at the deeper parts and
near regions of complex geology.

\subsubsection{Uncertainty due to noise, linearization errors, and
uncertain control
points}\label{uncertainty-due-to-noise-linearization-errors-and-uncertain-control-points}

\begin{figure*}
\centering
\includegraphics[width=0.700\hsize]{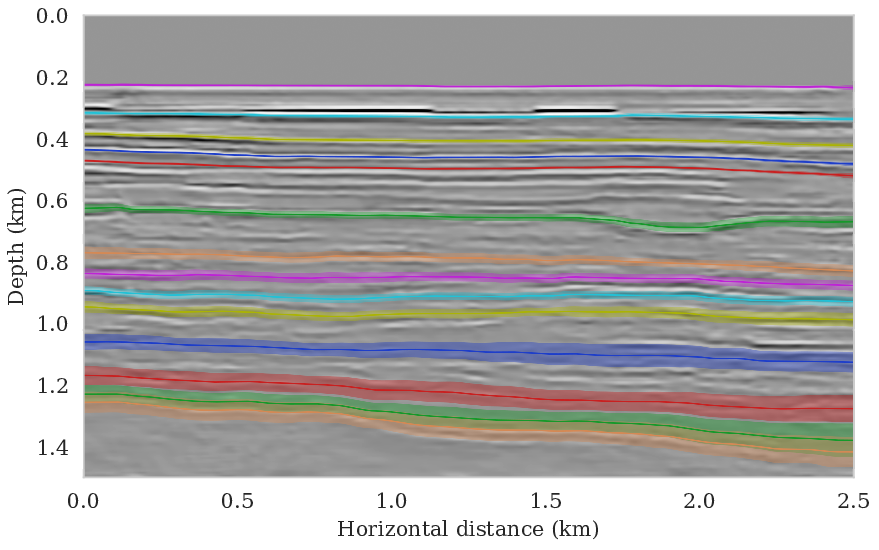}
\caption{Uncertainty in horizon tracking due to a combination of
uncertainties in imaging and control points
(equation~\ref{sum-horizons-points}).}\label{BG_horizon-uq-nondet}
\end{figure*}

Following the procedure described above, we also consider the effect of
randomness in the horizon tracking task itself. We mimic this by
imposing a distribution over the location of control points. As before,
we consider the case where we are equally confident in the location of
control points. The results for the conditional mean and the $99\%$
confidence interval are presented in Figure~\ref{BG_horizon-uq-nondet}.
As expected, uncertainties increase consistently with depth, close to
the boundaries, and in areas of complex geology. Compared to the
previous example, these effects are more pronounced, which we argue is
due to an increase in linearization errors at later times in the shot
data. This increase leads to more uncertainty in deeper sections.

\section{Discussion}\label{discussion}

The examples presented in this paper demonstrate the beneficial
regularization properties of deep priors as long as overfitting of noisy
data is avoided. Unfortunately, preventing overfit is challenging in
practice. To mitigate this issue, we proposed the use of conditional
mean estimates rather than maximum a posteriori (MAP) estimation even
though the former relies on sampling the posterior distribution, which
is computationally expensive. Based on our experience and other studies
\citep{Cheng_2019_CVPR, wilson2020bayesian}, estimates based on the
conditional mean are comparatively robust to overfitting and yield
superior results.

Even though having access to samples from the posterior has many
advantages, e.g., it gives us access to the conditional mean and
pointwise standard deviation estimates, its computational cost becomes
typically prohibitive for large dimensional problems with expensive
forward modeling operators. By using techniques from stochastic
optimization, we managed to partly offset these costs by avoiding exact
calculation of the multi-source data likelihood function. Similar to
stochastic optimization, sometimes employed to solve wave-equation based
optimization problems
\citep{vanLeeuwen2010IJGswi, haber2012effective, herrmann2012, lu2015, tu2015, li2018},
the gradient of the data misfit is calculated for artificially
constructed simultaneous source experiments. This reduces the number of
wave-equation solves for each gradient calculation significantly. In the
context of Langevin dynamics---the theory undergirding our Markov chain
Monte Carlo (MCMC) method to sample from the posterior---this stochastic
approximation corresponds to the stochastic gradient Langevin dynamics
(SGLD) method proposed by \citet{welling2011bayesian}. This approach, in
combination with a preconditioning scheme and stepsize schedule,
eliminates computationally prohibitive Metropolis-Hasting acceptance
steps. By means of several numerical experiments and empirical accuracy
measures, we established that the proposed SGLD algorithm is capable of
drawing samples from the posterior with a reasonable accuracy. Aside
from providing a reasonable assessment of the uncertainty, with
pointwise standard deviation increasing in complex areas or in areas of
relatively poor illumination, the samples from the posterior also
allowed us to propagate uncertainties due to errors in the shot data all
the way to the task of automatic horizon tracking or other tasks. With
very few exceptions, we are not aware of this type of work on relatively
large scale problems \citep{interrogation, CHO2021108333}.

While uncertainty quantification based on Bayesian inference certainly
has its merits, it comes at a significant computational price even for
2D problems. These computational costs are compounded by the fact that
the parameterization of seismic images in terms of a deep neural network
is highly overparameterized, making it more difficult to solve the
uncertainty quantification problem. Notwithstanding these challenges,
the use of deep priors has several distinct advantages. First, the
regularization comes from the inductive bias of the network architecture
itself, which is designed to favor natural images. Second, this approach
eliminates the need of having access to training data when compared to
method that really prior information encoded in pretrained networks.
Third, imposing a Gaussian prior on the networks weights is a common
regularization strategy \citep{krogh1992simple, Goodfellow-et-al-2016}.
Despite these advantages, the number of iterations needed by the SGLD
algorithm remains high and prohibitive for imaging problems in 3D. For
this reason, reducing the computational complexity of Bayesian inference
over the weights of deep prior networks remains an activate area of
research. For instance, \citet{izmailov2020subspace} proposes to project
the network weights onto a low-dimensional subspace and perform
posterior inference within this reduced subspace. Unfortunately,
construction of this reduced space can also be costly. For the purpose
of applying this work to 3D seismic imaging, one possibility is to
combine this dimensionality reduction approach with a different form of
CNNs that exhibit similar inductive biases to those that we used in this
work \citep{Lempitsky} but has fewer weights than the image
dimensionality \citep{heckel_deep_2019}. As a result, we may be able to
exploit the inductive bias of CNNs at a lower computational cost

Despite the fact that Bayesian inference based on MCMC methods, such as
SGLD, is well-studied and widely employed, it is challenging for
high-dimensional inverse problems that involve expensive forward
operator. Variational (Bayesian) inference
\citep{jordan1999introduction, rezende2015variational}, also known as
distribution learning, can potentially overcome these challenges. In
this approach, a neural network is trained to synthesize new samples
from a target distribution based on a collection of training samples.
Typically these samples are obtained by applying a series of learned
nonlinear functions to random realizations from a canonical
distribution. Early work on variational inference
\citep{rizzuti2020SEGpub, siahkoohi2020TRfuqf, zhang2020seismic, kothari2021trumpets, kovachki2021conditional, kruse2021hint, kumar2021SEGeuq, siahkoohi2021Seglbe, siahkoohi2020ABIpto, zhao2020bayesian}
shows encouraging results, which opens enticing new perspectives on
uncertainty quantification in the field of wave-equation based
inversion.

Uncertainty in solving (linear) inverse problems including seismic
imaging comes in two flavors. On the one hand, there is uncertainty
related to noise in the input (shot) data. On the other hand, there may
be modeling errors, such as linearization errors, which decrease with
the accuracy of the background velocity model. We plan to study how
these two types of uncertainty specifically affect the results in future
work, with a certain regard for the impact of the background velocity
model. Compared to the problem addressed in this paper, this would
entail multiple imaging experiments for different background velocity
models and is therefore more challenging. Recent developments in full
subsurface offset image volumes \citep{yang2020lrpo} and Fourier neural
operators \citep{li2021fourier} may prove essential in addressing this
problem.

\section{Conclusions}\label{conclusions}

Bayesian inference on high-dimensional inverse problems with
computationally expensive forward modeling operators, has been, and
continues to be a major challenge in the field of seismic imaging. Aside
from obvious computational challenges, the selection of effective priors
is problematic given the heterogeneity across geological scenarios and
scales exhibited by elastic properties of the Earth's subsurface. To
limit the possibly heavy-handed bias induced by a handcrafted prior, we
propose regularization via deep priors. During this type of
regularization, seismic images are restricted to the range of an
untrained convolutional neural network with a fixed input, randomly
initialized. Compared to conventional regularization, which tends to
bias solutions towards sometimes restrictive choices made in defining
these prior distributions, nonlinear deep priors derive regularizing
properties from their overparameterized network architecture. The
reparameterization of the seismic image by means of a deep prior leads
to a Bayesian formulation where the prior is a Gaussian distribution of
the weights of the network.

As long as overfitting can be avoided, regularization with deep priors
is known to produce perceptually accurate results, an observation we
confirmed in the context of controlled seismic imaging experiments.
Unfortunately, preventing fitting the noise is difficult in practice. In
addition, there is always the question how errors, e.g.,
bandwidth-limited noise or linearization errors, propagate to
uncertainty in the image and to certain tasks to be carried out on the
image, which for instance includes the task of automatic horizon
tracking. To answer this question and to avoid the issue of overfitting,
we propose to sample from the imaging posterior distribution and use the
samples to compute the conditional mean estimate, which in our
experiments exhibited more robustness to noise, and to obtain confidence
intervals for the tracked horizons via our probabilistic horizon
tracking framework.

Even though drawing samples from the posterior is computationally
burdensome, it allows us to mitigate the imprint of overfitting while it
is also conducive to a systematic framework mapping errors in shot data
to uncertainty on the image and task at hand. By means of two imaging
experiments derived from imaged seismic data volumes, we corroborated
findings in the literature that the conditional mean estimate, i.e., the
average over samples from the posterior distribution on the image, is
more robust to overfitting than the maximum a posteriori estimate. The
latter is the product of deterministic inversion. Aside from improving
the image quality itself with the conditional mean estimate, access to
samples from the posterior also allows us to compute pointwise standard
deviation on the image and confidence intervals on automatically tracked
horizons.

With few exceptions as of yet, no systematic attempts have been made to
account for uncertainties in the task of horizon tracking due to errors
in the seismic imaging itself. These errors are caused by noise,
linearization approximations, and uncertainty in the horizon tracking
process itself, the latter being possibly related to differences in the
selection of control points by different interpreters as part of the
task of automatic horizon tracking.

To validate the proposed probabilistic tasked imaging framework, we
considered realistic scenarios that are representative of two different
geological settings. Our findings include: empirical verification of the
accuracy of the samples from the posterior distribution; establishment
of the conditional mean as a robust estimate for the image; reasonable
estimates for the pointwise standard deviation on the image, showing an
expected increase in variability in complex geological areas and in
areas with poor illumination; and finally confidence intervals for the
automatic horizon tracking given the uncertainty on the image and errors
in the selection of control points guiding automatic horizon tracking.

\section{Related material}\label{related-material}

The SGLD iterations (equation~\ref{sgld}) require computing gradient of
the negative-log posterior with respect to the CNN weights. This
requires actions of the linearized Born scattering operator and its
adjoint. For maximal numerical performance, the just-in-time
\href{https://www.devitoproject.org/}{Devito}
\citep{devito-compiler, devito-api} compiler was used for the
wave-equation based simulations. To have access to the automatic
differentiation utilities of \href{https://pytorch.org/}{PyTorch}, we
expose Devito's matrix-free implementations for the migration operator
and its adjoint to PyTorch. In this way, we are able to compute the
gradients required by equation~\ref{sgld} with automatic differentiation
while exploiting Devito's highly optimized migration and demigration
operators. For the CNN architecture, we followed \citet{Lempitsky}. For
the automated horizon tracking we made use of
\href{https://github.com/xinwucwp/mhe}{software} written by
\citet{wu2018least}. For more details on our implementation, please
refer to our code on
\href{https://github.com/slimgroup/deep_inference_with_tasks}{GitHub}.

\section{Acknowledgments}\label{acknowledgments}

This research was carried out with the support of Georgia Research
Alliance and partners of the ML4Seismic Center.

\section{Appendix A}\label{appendix-a}

\subsection{Deep-prior network
architecture}\label{deep-prior-network-architecture}

The CNN architecture proposed by \citet{Lempitsky} is a variation of the
widely used U-net architecture, as described by
\citet{ronneberger2015u}. The U-net architecture is composed of an
encoder and a decoder module, where information, i.e., intermediate
values, from the encoder module is also passed to the decoder through
skip connections.

The following are the major differences between the deep prior
architecture \citep{Lempitsky} and the U-net architecture. Contrary to
U-net, the convolutional layers in the decoding module of the deep prior
architecture do not increase the dimensionality of the intermediate
values. Rather, dimensionality-preserving convolutional layers, that is,
stride-one convolutions, are augmented with user-defined interpolation
schemes to achieve upsampling. This enables the degree of smoothness in
the image space to be controlled by the interpolation kernel. Another
difference worth noting is the way intermediate values from the encoding
phase are incorporated into the decoding module through skip
connections. In the deep prior architecture, the intermediate values in
the encoding phase are passed through an additional convolutional layer
before being fed into the decoder module.

\begin{figure*}
\centering
\includegraphics[width=1.000\hsize]{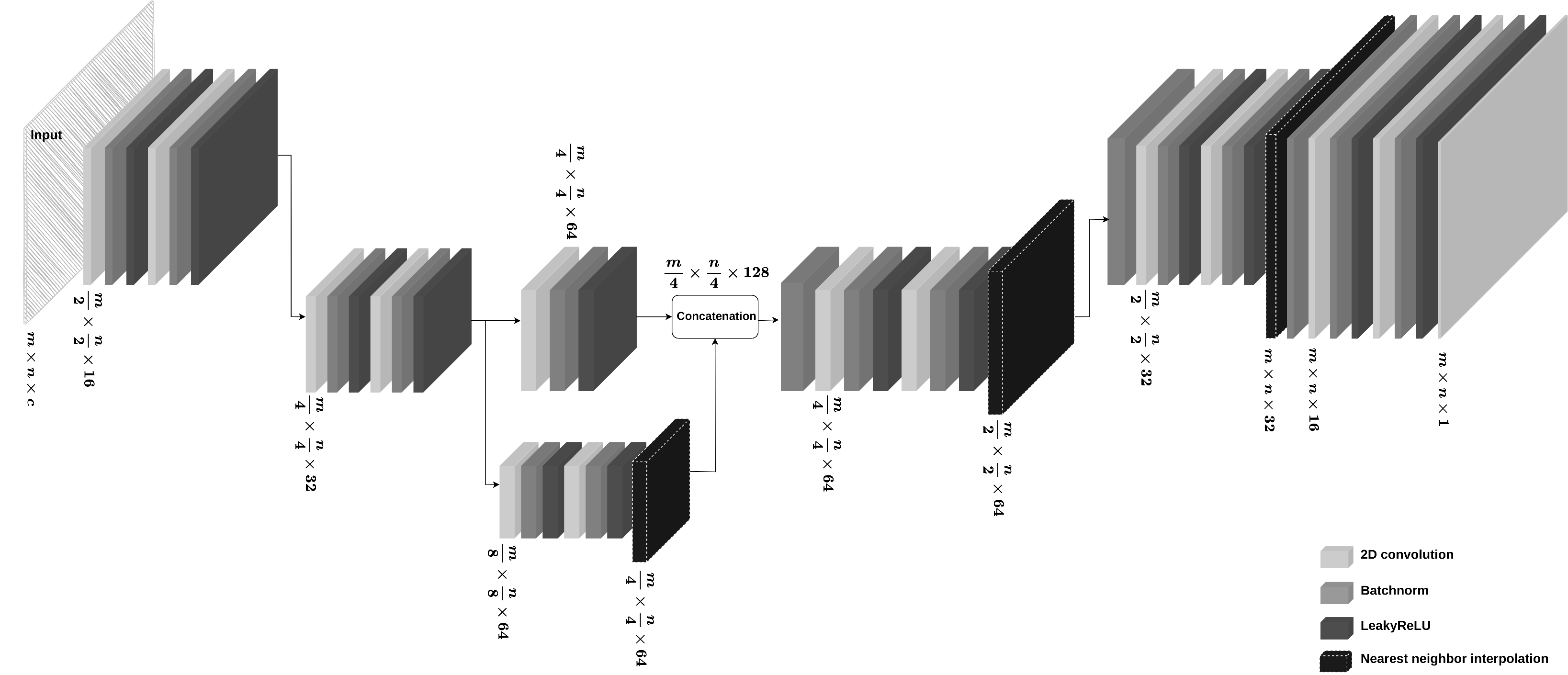}
\caption{The architecture used for the Parihaka
example.}\label{architecture_deep_prior}
\end{figure*}

Figure~\ref{architecture_deep_prior} illustrates the exact deep prior
architecture used in the Parihaka example in this paper, which closely
follows the architecture advocated by \citet{Lempitsky}. The blocks in
the figure represent the intermediate values in the network, where the
color indicates the operation that produced these values. For instance,
the lightest gray blocks are intermediate values obtained from applying
two-dimensional convolutions, whereas the darkest gray block with dashed
white edges represents the result of a user-selected interpolation
method, which is in this instance nearest neighbor interpolation. For
further information regarding the rest of the colors, please refer to
the legend in Figure~\ref{architecture_deep_prior}. The left most shaded
block in Figure~\ref{architecture_deep_prior} is the
$m \times n \times c$ input, which in the case of deep priors is a fixed
random array, with $m,\,n$ indicating its horizontal and vertical
dimensions, and $c$ which represents the number of channels. If a layer
alters the dimension of an intermediate value, the new dimensions are
inserted adjacent to the representative block. For instance, the
leftmost convolutional layer that takes in the input noise changes its
dimensionality to $\frac{m}{2} \times \frac{n}{2} \times 16$. All
two-dimensional convolutional layers use $5 \times 5$ kernels. The
convolutional layers that reduce the horizontal and vertical dimensions
have a stride of two while all the other convolutional layers have a
stride of one. Lastly, we apply a fixed scaler to the output of the
network such that the range of the output values for the CNN with fixed
input and randomly drawn weights
$\B{w} \sim \mathrm{N} \big (\B{w} \mid \B{0}, \lambda^{-2}\B{I} \big )$
covers the a priori known range of amplitudes in the unknown
perturbation model.

The Compass example uses the same architecture as
Figure~\ref{architecture_deep_prior}, except that it includes an
additional downsampling layer in the encoder and an additional
upsampling layer in the decoder. We found the addition of these layers
to be helpful since the frequency content of the model in the Compass
example was higher and the dimensions larger.

\bibliography{paper}

\end{document}